\documentclass[pra,amssymb,amsmath,superscriptaddress,twocolumn,footinbib]{revtex4-2}
\usepackage{braket}
\usepackage{float}
\usepackage{accents}
\usepackage{tikz}
\usepackage{multirow}
\usepackage{booktabs}
\usepackage[export]{adjustbox}
\usepackage{graphicx}
\usepackage{afterpage}
\usepackage[page,toc]{appendix}
\usepackage{amsthm}
\usepackage{dsfont}
\usepackage{hyperref}
\usepackage{chngcntr}
\hypersetup{
	colorlinks=true,
	linkcolor=blue,
	filecolor=magenta,      
	urlcolor=magenta,
	citecolor=teal
}
\usepackage{xcolor}
\usepackage[normalem]{ulem}
\usepackage{soul}
\newcommand{\effcnot}{C_X^{M\rightarrow G}}
\DeclareMathOperator{\Tr}{Tr}

\newcommand{\outprod}[1]{\ket{#1}\!\!\bra{#1}}

\newcommand{\proj}[2]{\ket{#1}\!\!\bra{#2}}
\newcommand{\crt}[2][]{\hat{#2}^\dagger_{#1}}
\newcommand{\anh}[2][]{\hat{#2 }_{#1}}

\usepackage{afterpage}

\usepackage{dcolumn}% Align table columns on decimal point
\usepackage{bm}
\usepackage{blkarray}
\usepackage{xparse}
\usepackage{mathtools}

\usepackage[shortlabels]{enumitem}

\usepackage{xcolor}

\begin{document}
	\preprint{APS/123-QED}
	%opening
	\title{Interfacing Gottesman-Kitaev-Preskill Qubits to Quantum Memories}
	
	\author{Prajit Dhara}%
	\email[]{prajitd@arizona.edu}
	\affiliation{Wyant College of Optical Sciences, The University of Arizona, Tucson, AZ 85721}
	\affiliation{NSF-ERC Center for Quantum Networks, The University of Arizona, Tucson, AZ 85721}
	
	\author{Liang Jiang}
        \affiliation{NSF-ERC Center for Quantum Networks, The University of Arizona, Tucson, AZ 85721}
	\affiliation{Pritzker School of Molecular Engineering, University of Chicago, Chicago, IL 60637}
	
	\author{Saikat Guha}
		\email[]{saikat@umd.edu}
	\affiliation{Wyant College of Optical Sciences, The University of Arizona, Tucson, AZ 85721}
	\affiliation{NSF-ERC Center for Quantum Networks, The University of Arizona, Tucson, AZ 85721}
         \affiliation{Department of Electrical and Computer Engineering, University of Maryland, College Park, MD 20742}
	
	%\date{Started: June 18, 2022 \quad Current: \today }

	\begin{abstract}
		Gottesman-Kitaev-Preskill (GKP) states have been demonstrated to pose significant advantages when utilized for fault-tolerant all optical continuous-variable quantum computing as well as for  quantum communications links for entanglement distribution. However interfacing these systems to long-lived solid-state quantum memories has remained an open problem. Here we propose an interface between quantum memories and GKP qubit states based on a cavity-mediated controlled displacement gate. We characterize the quality of memory-GKP entanglement as a function of cavity parameters suggesting optimal regimes of operation for high-quality state transfer between either qubit states. We further extend this protocol to demonstrate the creation of GKP cluster states by avoiding the requirement of ancillary optical quadrature-squeezed light. Utilizing post-selected entanglement swapping operations for GKP qubits, we demonstrate the utility of our protocol for high-rate entanglement generation between quantum memories. Extensions and derivatives of our proposal could enable a wide variety of applications by utilizing the operational trade-offs for qubits encoded in memory and in the GKP basis.
	\end{abstract}
	
	\maketitle

	\section{Introduction}
	\label{sec:intro}
	
	The broad field of quantum information processing is heavily reliant on the use of bosonic systems at its core. Qubit encodings of a general bosonic mode, particularly photonic modes, are key to many applications in this domain. Of the large class of qubit encodings, the most popular are the discrete variable (DV) encodings, where the presence/absence of a photon or the presence of a single photon in a pair of orthogonal modes encodes the qubit.  An emerging class of encodings are the so-called continuous variable (CV) encodings where the qubit levels are determined by orthogonal phase space distributions. The interest in these types of qubits has been bolstered by the relative ease of logical quantum operations (i.e. linear optics, assisted by photon number resolved detection) for information processing. The most famous CV qubit encoding is the Gottesman-Kitaev-Preskill (GKP) encoding~\cite{Gottesman2001-hd} -- the qubit states are defined as the simultaneous eigenstates of displacement operations along two quadrature directions. The lattice-like phase space distribution is the signature of the GKP encoding -- it is particularly suitable for the detection and correction of random displacement errors~\cite{Gottesman2001-hd,Harrington2001-uy} and excitation loss errors~\cite{Albert2018-qr,Noh2019-gd}. Proposals have been made for the application of the GKP qubit encoding for tasks in quantum computing~\cite{Fluhmann2019-sj,Campagne-Ibarcq2020-az} and long-distance quantum communications based on repeaters~\cite{Fukui2021-ay,Rozpedek2021-nk,Rozpedek2023-vi,Schmidt2024-ut}. However a multitude of these proposals implicitly rely on some faithful, and error-suppressed technique to `store' the qubit.
	
	Although methods for optical path delay (for e.g. fiber delay lines)  would serve as effective memories for photonic GKP qubits, the storage time is limited by photon-loss induced random displacement of the qubit -- this inherently imposes a hard cutoff on the memory storage times. Quantum memories that can encode the qubit information in an auxiliary quantum system (ideally with additional error detection and mitigation methods)  are more practical for the purposes of reliable and faithful storage. Many proposals for these memories have been put forward~\cite{Chen2021-do,Bhaskar2020-kv,Nguyen2019-hg,Sukachev2017-kg,Schupp2021-of} -- each candidate has their own set of merits and limitations. Interfacing photonic qubits with memories usually require some additional interaction. For DV encodings, the broad class of memory-photon interfaces are achieved via direct interactions~\cite{Duan2004-qs,Chen2021-do,Bhaskar2020-kv},, or by photonic entanglement swap assisted logical multi-qubit operations~\cite{Barrett2005-sm,Dhara2023-zv}. However for CV encodings such interactions are limited to specific use-cases; a general methodology is not well defined.
	
	In this work, we propose a protocol for the implementation of a hybrid two-qubit gate on a memory qubit and a GKP qubit. Our protocol relies on an atomic state dependent photonic phase rotation operation to implement a controlled-displacement of the GKP qubit. Since displacements are logical Pauli operations in the GKP encoding, the gate action can be an effective CNOT or CPHASE gate where the memory qubit acts as the control. We highlight the details of our protocol and perform a detailed analysis of the associated interactions as a function of interaction, cavity and state parameters. With the aid of this entangling operation, we propose and analyze a series of applications, namely,
	\begin{enumerate}
		\item Quantum circuits for bi-directional quantum state transfer via qubit teleportation.
		\item Extended protocols for the creation of larger GKP qubit cluster states
		\item Near-deterministic memory-memory entanglement generation using post-selected dual-homodyne based GKP basis Bell projection.
	\end{enumerate}
	Our protocol for the hybrid qubit domain entangling operation promises multiple applications for associated quantum information processing tasks. We motivate and suggest a roadmap of potential ideas for future studies to explore.
	
	%quantum circuits for bi-directional quantum state transfer between either qubits. We extend our protocol for the creation of GKP qubit cluster states -- we discuss alternative strategies for the generation of arbitrarily sized cluster states. Additionally by leveraging near-deterministic post-selected dual-homodyne based entanglement swaps for GKP qubits we show how GKP-spin entangled qubits generated by our protocol could be used for high efficiency memory-memory entanglement generation over low-loss channels surpassing the performance of common DV qubit encodings. 
	
	The article is organized as follows -- we provide the relevant background for the GKP qubit encoding, along with the interaction mechanism of bosonic modes with the atom-cavity system in Sec.~\ref{sec:background_system_setup}. The system description and setup introduces the systems under consideration and the interactions of the various sub-components with each other; we develop the input-output coupling relation as function the various system parameters. We propose and discuss our protocol in detail in Sec.~\ref{sec:interaction_details}; we emphasize the requirements on the underlying interactions for the protocol to work in the desired manner. Additionally we develop a circuit model for the entire effective two-qubit interaction. Section~\ref{sec:bidirectional_transfer} covers the circuits for bi-directional quantum state teleportation as well as the final state quality after heralding using imperfect measurements in either qubit domain. We extend our protocol for the generation of GKP qubit cluster states in Sec.~\ref{sec:cluster_generation}; we discuss few alternate approaches and their limitations. Section~\ref{sec:gkp_assisted_swap} analyzes the performance of GKP qubit assisted entanglement generation between memories using post-selected dual-homodyne measurements for entanglement swapping. We conclude our article with a roadmap for potential future studies that may be pursued to extend the utility of the proposed protocol.

	\section{Background and System Setup}
	\label{sec:background_system_setup}
	
	\subsection{GKP Qubits: Definitions and Operations}
	In this article, we work with $\hbar=1$ units. Consider a single bosonic mode with the creation ($\hat{a}^\dagger$) and annihilation ($\hat{a}$) operators defined in terms of the canonical position ($\hat{q}$) and momentum ($\hat{p}$) as $\hat{a}=(\hat{q}+i \hat{p})/\sqrt{2}; \hat{a}^\dagger=(\hat{q}-i \hat{p})/\sqrt{2}$. The displacement of states along the $\hat{q}$ and $\hat{p}$ quadratures are defined as $\hat{X}(u)=e^{-iu\hat{p}}$ and $\hat{Z}(v)=e^{iv\hat{q}}$ respectively. The logical square-lattice GKP qubit subspace~\cite{Gottesman2001-hd} is defined as the eigenspace formed by  $\hat{X}(2\sqrt{\pi})$ and $\hat{Z}(2\sqrt{\pi})$ displacement operators. The logical qubit states can then be be expressed (in terms of position eigenstates) as,
	\begin{align}
		\begin{split}
			\ket{\bar{0}}_G &= \sum_{t=-\infty}^{\infty} \ket{2t\sqrt{\pi}}_q;\\
			\ket{\bar{1}}_G &= \sum_{t=-\infty}^{\infty} \ket{(2t+1)\sqrt{\pi}}_q.
		\end{split}
	\end{align}
	As is apparent from the definitions, the logical $	\ket{\bar{0}}_G $($	\ket{\bar{1}}_G$) state is a superposition of Dirac delta peaks at even (odd) positions of a square lattice of $\sqrt{\pi}$-spacing in phase space. The computational basis states $\ket{\bar{\pm}}_G$ are equivalently defined in terms of \emph{momentum} eigenstates.The logical Pauli operations for the GKP basis are defined as $\hat{X}\equiv\hat{X}(\sqrt{\pi}); \hat{Z}\equiv\hat{Z}(\sqrt{\pi})$ and $\hat{H}\equiv R(\pi/2)$ -- they consitute of phase space displacements and rotations. Other variants of the GKP codes are possible with a modified lattice definition~\cite{Albert2018-qr,Noh2019-gd}; however they are all equivalent upto a symplectic transformation of the canonical coordinates. 
	
	Owing to the unphysical nature of the logical GKP qubits (i.e.\ infinite superposition of infinite energy states), we adopt a physical GKP qubit definition. The physical qubit states have finite mean photon number (i.e.\ finite energy) and are properly normalized. In the position and momentum representation the states are defined as 
	\begin{align}
		\begin{split}
			\ket{\tilde {L}}_{G}\propto & \sum_{t=-\infty}^{\infty} e^{-\pi\sigma_2^2(2t+L)^2/2}\\
			& \qquad \times \int \exp\left[-\frac{(q-(2t+L)\sqrt{\pi})^2}{2\sigma_1^2}\right]\ket{q} dq\, ,
		\end{split}
	\end{align}
	where $ L=\{0,1\}$. The variance parameters $\sigma_1,\sigma_2$ can be unique and independent. In the given representation $\sigma_1$ controls the peak width (which ensures the mean photon number is finite) and $\sigma_2$ controls the width of the overall Gaussian envelope (which ensures the superposition is finite). In the limit $\sigma_1,\sigma_2\rightarrow0$, we retrieve the ideal GKP qubit states $\ket{L}_G$. The quadrature variances are related to the peak width and envelope parameters -- we shall operate with the assumption $\langle \hat{q}^2\rangle =\langle \hat{p}^2 \rangle \equiv\sigma^2_{q/p} =\sigma^2_1/2=\sigma^2_2/2$. The logical qubit gates for the physical qubits remains the same; however the qubit states are not orthogonal for all $\sigma_1=\sigma_2=\sigma$. 
	
	An alternate definition of the GKP qubit ignores the phase space envelope~\cite{Menicucci2014-wo} -- it is convenient to treat them as randomly displaced versions of the ideal qubit states. This form of the state is defined by 
	\begin{align}
		\begin{split}
			\ket{\accentset{\approx}{L}}_{G}\propto \int du \, dv \, \Gamma(u,v) \, \hat{X}(u) \hat{Z}(v)  \ket{\bar{L}}_G
		\end{split}
	\end{align}
	where $\Gamma(u,v)$ is a bi-variate Gaussian function that controls the extent of displacement,
	\begin{align}
		\Gamma (u,v) = \frac{1}{2\pi {\sigma_u \sigma_v} }\exp\left[-\frac{u^2}{2\sigma_u^2} -\frac{v^2}{2\sigma_v^2} \right].
	\end{align}
	Although this definition is still un-physical (due to the infinite superposition), operations on the qubit state can be effectively treated as a evolution of the displacement variances (i.e. $\sigma_u,\sigma_v$). Typically, to maintain symmetry, we set equal variances for both quadratures $\sigma_u= \sigma_v=\sigma$.
	
	The various alternative descriptions of the GKP state can be shown to be equivalent to each other~\cite{Matsuura2020-qv}. Readers may look at Appendix~\ref{app:gkp_definition} for additional background material.
	%{\color{red} \noindent PD: Does this section need any additional details?}
	
	\subsection{Atom-Cavity System Details}
	\label{sec:setup_atom_cav}
	
	\begin{figure}[]
		\centering
		\includegraphics[width=0.85\linewidth]{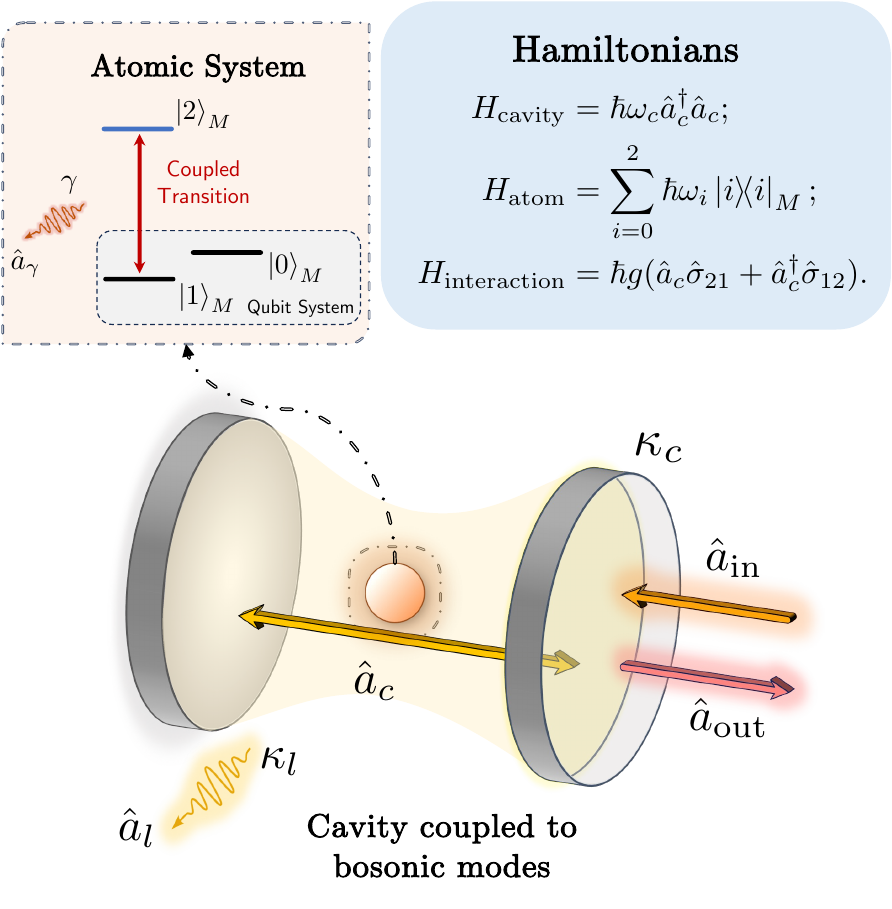}
		\caption{Overview of the cavity system coupled to the atomic 3-level system. The relevant modes, eigen levels and parameters are labelled. The specific system Hamiltonians for the cavity~$(H_\mathrm{cavity})$ and the atom $(H_{\mathrm{atom}})$  as well as the interaction Hamiltonians $(H_\mathrm{interaction})$ are highlighted. Detailed descriptions can be found in Sec.~\ref{sec:setup_atom_cav}.
		} 
		\label{fig:cavity_atom_system}
	\end{figure}
	In this article, ee will focus on the interaction of photons in bosonic modes with a single 3-level atom coupled to a single-sided cavity; relevant parameters are defined subsequently. 
	
	The atomic system (which we shall henceforth refer to as the `memory')  is assumed to support three level $L$-system model, with a single coupled interaction at optical frequencies. The energy eigenlevels of the atom are denoted by $ \ket{0}_M,\ket{1}_M $ and $\ket{2}_M  $  with eigenenergies $ \hbar\omega_0,\hbar\omega_1,\hbar\omega_2 $ respectively. The $ \ket{1}_M\leftrightarrow\ket{2}_M $ transition is the transition of interest, and we define the logical qubit in terms of the $ \ket{0}_M,\ket{1}_M $ states.  We account for atomic decay from $\ket{2}_M\!\rightarrow\!\ket{1}_M $  at a rate denoted by  $\gamma_m$ where photons are emitted into a (unrecovered) decay mode $ \hat{a}_\gamma $ (other than the coupled cavity mode). Additionally, we assume that the $ \ket{2}_M\!\rightarrow\!\ket{0}_M$ decay is forbidden by selection rules, and the $ \ket{0}_M\!\rightarrow\!\ket{1}_M $ decay is suppressed (or is insignificant in the time scales of our interaction). We define the atomic population ($i=j$) and coherence ($i\neq j$) operators  as $\hat{\sigma}_{ij}=\proj{i}{j}_M$. 
	
	The interaction of bosonic modes interacting with the single-sided cavity is central to the implementation of our protocol. The incoming, outgoing and cavity modes' field operators are respectively labeled as $ \hat{a}_\mathrm{in}, \hat{a}_\mathrm{out} \text{ and } \hat{a}_c$. The field operators are coupled by the standard input-output relation $\hat{a}_\mathrm{out}=\sqrt{\kappa_c}\hat{a}_c+\hat{a}_\mathrm{in}$
	with cavity-coupling rate of $ \kappa_c $. Photons may be lost from the cavity into an auxiliary mode $ \hat{a}_l $ with a loss rate denoted by $ \kappa_l $.  We call $\kappa=\kappa_c+\kappa_l$ the total coupling rate.
	
	The cavity, atom, and interaction Hamiltonians of the joint system is given by~\cite{Walls2012-fo,Gardiner1985-hn},
	\begin{align}
		\begin{split}
			H_{\mathrm{cav.}}&=\hbar \omega_c \hat{a}^\dagger_c\hat{a}_c;\\
			H_{\mathrm{atom}}&= \sum_{i=0}^{2} \hbar \omega_i \outprod{i}_M;\\
			H_{\mathrm{int.}}&=\hbar g(\hat{a}_c \hat{\sigma}_{21} + \hat{a}^\dagger_c\hat{\sigma}_{12}).
		\end{split}
	\end{align}
	$ \omega_c $ is the cavity resonance frequency, $\omega_i$ is the eigenfrequency for level $\ket{i}_M$ and $ g $ is the cavity mode-atom coupling strength. The complete Hamiltonian is then $\hat{H}={H_{\mathrm{cav.}}}+{H_{\mathrm{atom}}}+{H_{\mathrm{int.}}} $.
	
	With the aid of the Heisenberg-Langevin formulation~\cite{Gardiner1985-hn,Walls2012-fo} for the time dynamics of the atomic and bosonic mode operators, we may analyze the evolution of the state of incoming mode after interaction with the coupled atom-cavity system. The following Langevin equations can be derived by considering interaction of the atomic/cavity modes with their corresponding reservoir modes --
	\begin{subequations}
		\begin{align}
			&	\dot{\hat{a}}_c=\frac{-i}{\hbar} \left[\hat{a}_c,\hat{H}\right]-\sqrt{\kappa_c}\hat{a}_\mathrm{out}-\sqrt{\kappa_l}\hat{a}_l+\frac{\kappa}{2}\hat{a}_c ;\\
			& \dot{\hat{\sigma}}_{12}=\frac{-i}{\hbar} \left[\hat{\sigma}_{12},\hat{H}\right]
			+(\hat{\sigma}_{22}-\hat{\sigma}_{11}) \left(-\frac{\gamma_m}{2}\hat{\sigma}_{12} + \sqrt{\gamma_m} \hat{a}_\gamma \right).
		\end{align}
	\end{subequations}
	Simplifying these expressions by evaluating the operator commutators with the Hamiltonian $[\,\cdot\,,\hat{H}]$ and performing the operator Fourier transform, yields the following spectral mode transformation relations
	\begin{subequations}
		\begin{align}
			-i\omega {\hat{a}}_c (\omega)&= -i\omega_c\hat{a}_c(\omega)-ig\hat{\sigma}_{12} \nonumber\\
			&\qquad-\sqrt{\kappa_c}\hat{a}_\mathrm{out}(\omega)-\sqrt{\kappa_{l}}\hat{a}_l(\omega)+\frac{\kappa}{2}\hat{a}_c(\omega);\\
			-i\omega {\hat{\sigma}}_-&=-i\omega_a\hat{\sigma}_{12} + ig\hat{a}_c (\omega) \hat{\sigma}_z -\frac{\gamma_m}{2}\hat{\sigma}_z \hat{\sigma}_{12} \nonumber \\
			&\qquad+ \sqrt{\gamma_m} \hat{a}_\gamma (\omega) \hat{\sigma}_z,
		\end{align}
	\end{subequations}
	where $\hat{\sigma}_z=\hat{\sigma}_{22}-\hat{\sigma}_{11}$. Making the assumption that $\langle\hat{\sigma}_z\rangle\approx-1$ (i.e.\ the excited state $\ket{2}_M$ is only virtually populated for large detunings $\omega_a - \omega \gg g{\braket{\hat{a}_c^\dagger \hat{a}_c} }^{1/2}$), one can show that the incoming and outgoing modes are coupled via the relation,
	\begin{align}
		\hat{a}_{\text {in }}(\omega)
		&=r(\omega) \hat{a}_{\text {out }}(\omega)+l_C(\omega) \hat{a}_{l}(\omega)+ l_A(\omega) \hat{a}_{\gamma}(\omega)
	\end{align} 
	where
	\begin{subequations}
		\begin{align}
			&r(\omega)=1- \frac{ \kappa_{c}}{\left(\kappa/2-i \Delta_{c}+\frac{g^{2}}{\gamma_m/2-i \Delta_{a}}\right)}; \\
			&l_C(\omega)=-\frac{\sqrt{\kappa_l \kappa_c}}{\left(\kappa/2-i \Delta_{c}+\frac{g^{2}}{\gamma_m/2-i \Delta_{a}}\right)}; \\
			&l_A(\omega)=\frac{{-ig\sqrt{\gamma_m \kappa_c}}/{(\gamma_m/2-i\Delta_a)}}{\left(\kappa/2-i \Delta_{c}+\frac{g^{2}}{\gamma_m/2-i \Delta_{a}}\right)}.
		\end{align}
	\end{subequations}
	with the substitutions $ \Delta_c=\omega_c-\omega; \Delta_a=\omega_a-\omega$ ( see Appendix~\ref{app:cavity_atom_langevin} for detailed derivations of these expressions). The mode transformation coefficients maybe re-expressed in terms of the atom-cavity cooperativity $C=4g^2/(\kappa\gamma_m)$ and coupling efficiency $\zeta=\kappa_c/(\kappa_c+\kappa_l)$ as 
	\begin{subequations}
		\begin{align}
			& r(\omega)=1- \frac{2\zeta}{\left(1-2i{\Delta_c}/{\kappa}+{C}/{(1-2i \Delta_{a}/\gamma_m)}\right)} \\
			& l_C(\omega)=-\frac{2\sqrt{\zeta\,(1-\zeta)}}{\left(1-2i {\Delta_c}/{\kappa}+{C}/{(1-2i \Delta_{a}/\gamma_m)}\right)} \\
			& l_A(\omega)=\frac{{-2i\sqrt{\zeta C}/(1-2i\Delta_a/\gamma_m)}}{\left(1-2i {\Delta_c}/{\kappa}+{C}/{(1-2i \Delta_{a}/\gamma_m)}\right)}.
		\end{align}
	\end{subequations}
	In any interaction, modes represented by $\hat{a}_\gamma $ and $\hat{a}_l$ will be lost since they represent scattered field operators - hence it is appropriate to call $r(\omega)$ the reflectivity, $l_C(\omega)$ the cavity loss coefficient and $l_A(\omega)$ the atomic loss coefficient.  
	
	The input-output relations have been similarly derived in previous works~\cite{Duan2004-qs,Hacker2019-ep,Hastrup2022-wy,Reiserer2015-wt,Weigand2018-lb} to demonstrate conditional phase rotations assisted CPHASE gates for dual rail photonic qubits~\cite{Duan2004-qs} and on coherent pulses to generate Schrodinger cat states of light~\cite{Weigand2018-lb,Hacker2019-ep}. Our proposal requires the implementation of a cavity resonant interaction $(\Delta_c=0)$ with a high cooperativity $(C\gg1)$ inducing the strongly coupled interactions. Given an input pulse shape $f(t)$, with a temporal bandwidth of $\tau_f$, the scheme is successfully implemented when the pulse is \emph{slowly varying compared to the cavity coupling rate} $\kappa\gg |\partial_tf(t)/f(t)|$ and the temporal bandwidth satisfies $\tau\gg1/\kappa$~\cite{Duan2004-qs}. Without these assumptions the input-output coupling relation must account for photon number induced non-linearities (for e.g. photon-blockade) to accurately model the system~\cite{Duan2004-qs,Kiilerich2020-xu}.
	
	The memory dependent phase rotation on the incoming state can be analyzed by considering the special case where $C\gg1$ and $ \kappa=\kappa_c; \kappa_l\rightarrow0 \Rightarrow\zeta=1 $. If the atomic state is $ \ket{1}_m $, the cavity is strongly coupled to the transition -- thus the cavity reflection coefficient is  $ r^{(1)}\approx1$ with loss becoming negligible i.e.\ $ l_C^{(1)}=l_A^{(1)}\approx0 $. However for the atomic state $\ket{0}_M$, the lack of any coupled transition yields a zero cooperativity. This in turn drives the cavity reflection coefficient to $ r^{(0)}=-1$ yielding the relative phase difference.

	\section{Hybrid Entangling Interaction Between GKP Qubits and Memories}
	\label{sec:interaction_details}
	\begin{figure*}
		\centering
		\includegraphics[width=\textwidth]{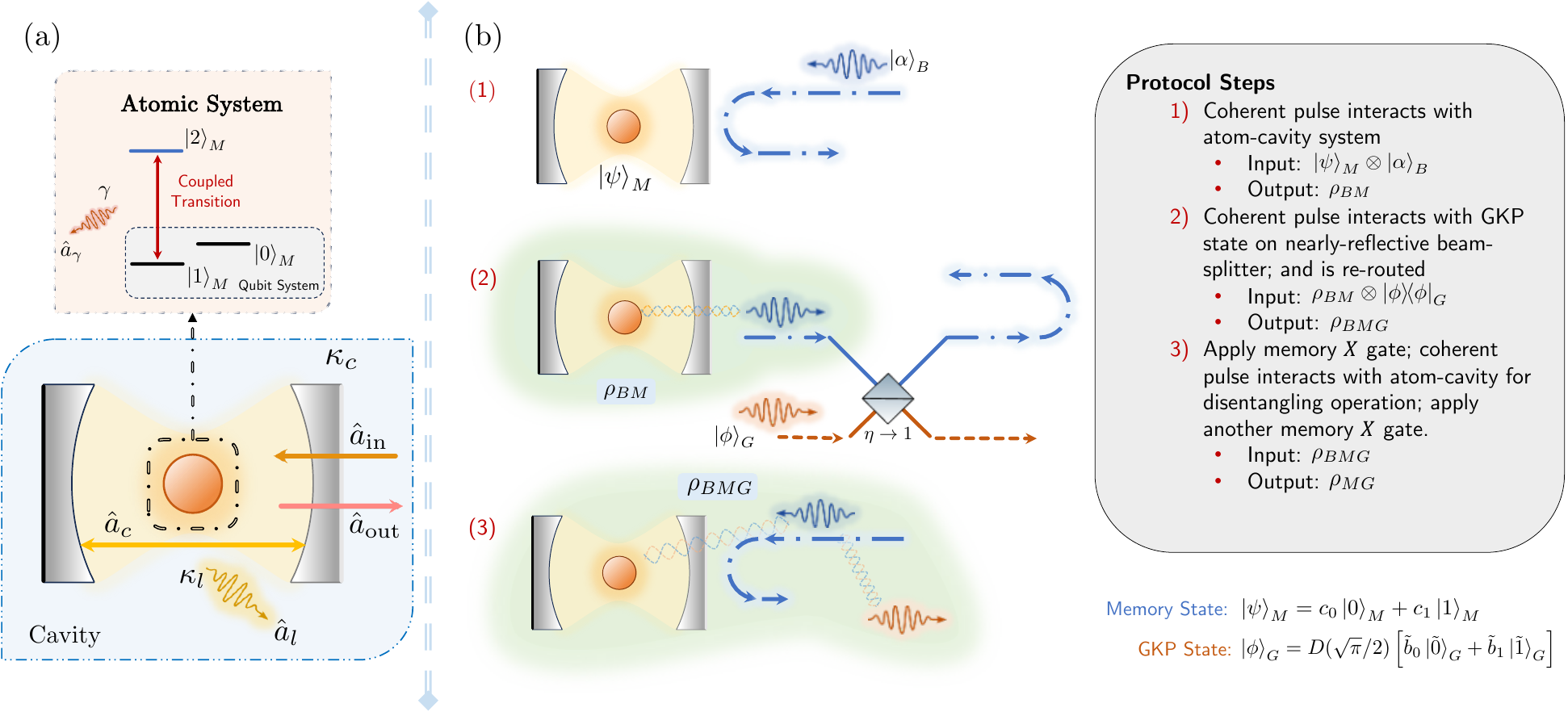}
		\caption{Overview of the interaction of the atomic system with the GKP qubit for the effective $\effcnot$ gate -- (a) Details of the cavity (blue bounding box) coupled to the atomic 3-level system (orange bounding box). The relevant modes, eigenlevels and parameters are labeled; details can be found in Sec.~\ref{sec:background_system_setup}. (b) Overview of the protocol implementing the $\effcnot$ gate. Detailed descriptions of the assumptions can be found in Sec.~\ref{sec:interaction_details}. In each step, we depict the entangled subsystems with a green halo in the background., as well as twisted line pairs.  Not shown explicitly are the Pauli gates applied on the atomic system, and the gates involved in the preparation of both the atomic and GKP states.
		} 
		\label{fig:interaction_cav}
	\end{figure*}
	We rely on the memory state-dependent phase rotation of the reflected photonic pulses to implement an entangling operation on a bosonic mode initialized as a GKP qubit state. Here-on we use the subscripts $B, M$ and $G $ for the bosonic mode, memory qubit and GKP qubit (i.e.\ the bosonic mode populated by the qubit) respectively. We begin by considering the interaction of a coherent pulse $\ket{\alpha}_B$ excited in the spectral pulse shape $f(\omega)$ -- additionally for ease of  analysis, we introduce the following equivalent notation for the coherent pulse to track the spectral response of the atomic system,
	\begin{align}
		\ket{\alpha}_B
		&\equiv \bigotimes_{\omega} \ket{\alpha f(\omega) d\omega}_B
	\end{align}
	The interaction of the pulse with an arbitrary memory state $\ket{\psi}_M=c_0\ket{0}_M+c_1 \ket{1}_M$ where $c_0,c_1\in \mathbb{C}$ and $|c_0|^2+|c_1|^2=1$ yields the output state (with all loss modes traced out),
	\begin{align}
		\begin{split}
			\rho_{BM}=&|c_0|^2 \outprod{\alpha_0}_B\otimes \outprod{0}_M \\
			&+ |c_1|^2\outprod{\alpha_1}_B\otimes \outprod{1}_M \\
			&+c_0 c_1^*\lambda \proj{\alpha_0}{\alpha_1}_B\otimes \proj{0}{1}_M \\
			&+c_0^* c_1 \lambda^* \proj{\alpha_1}{\alpha_0}_B\otimes \proj{1}{0}_M
			\label{eq:spin_coh_st}
		\end{split}
	\end{align}
	where $\lambda= \braket{l_C^{(1)}(\omega)\alpha f(\omega)|l_C^{(0)} (\omega)\alpha f(\omega)}\times \braket{l_A^{(1)}(\omega)\alpha f(\omega)|0}$ is the effective joint state dephasing factor. The reflected bosonic mode has a memory state dependent amplitude $	\alpha_0 = r^{(0)*}(\omega)\alpha f(\omega)$ and $\alpha_1 =r^{(1)*}(\omega)\alpha f(\omega)$. For $ C\gg1$ and $\Delta_c\rightarrow0$, we have the approximate relation $r^{(0)*}(\omega)\approx -r^{(1)*}(\omega)$ and $\lambda\rightarrow1$. Readers may note that at this stage the memory is entangled with the coherent pulse; by applying a Hadamard gate on the memory followed by a $\hat{Z}$ basis measurement once generate the Schrodinger cat states. This method of state generation has already been demonstrated in Ref.~\cite{Hacker2019-ep} and an extension of this interaction (with the input mode possibly excited in a squeezed state instead of a coherent pulse) was proposed for the generation of GKP qubits in Refs.~\cite{Weigand2018-lb,Hastrup2022-wy}. 
	
	For the purposes of the present article, the entangling interaction is achieved by having the reflected bosonic mode interact with an arbitrary pre-displaced GKP qubit state on a nearly reflective beamsplitter ($\eta\rightarrow1; \theta = 0$). Our beamsplitter convention for input modes $\hat{a}_1,\hat{a}_2$ and output modes $\hat{a}'_1,\hat{a}'_2$ is 
	\begin{align}
		\begin{pmatrix}
			\hat{a}'_1\\
			\hat{a}'_2
		\end{pmatrix}=\begin{pmatrix}
			\sqrt{\eta} & e^{i\theta}\sqrt{1-\eta}\\
			-e^{-i\theta}\sqrt{1-\eta} & \sqrt{\eta}
		\end{pmatrix} \begin{pmatrix}
			\hat{a}_1\\
			\hat{a}_2
		\end{pmatrix}
	\end{align}
	The input GKP state is considered to be an arbitrary pre-displaced qubit of the form $\ket{\phi}_G= D(\sqrt{\pi}/2)\left[\tilde{b}_0 \ket{\tilde{0}}_G+ \tilde{b}_1 \ket{\tilde{1}}_G\right] $ with  $\tilde{b}_0,\tilde{b}_1\in \mathbb{C}$ and $|\tilde{b}_0|^2+|\tilde{b}_1|^2=1$. After the beamsplitter interaction which approximates a displacement on the GKP state, the joint state is given by 
	\begin{widetext}
		\begin{align}
			\begin{split}
				\rho_{BMG}\approx & |c_0|^2 \outprod{\alpha_1}_B\otimes \outprod{0}_M \otimes  D(\alpha_0\sqrt{1-\eta})\rho_G D^\dagger(\alpha_0\sqrt{1-\eta}) \\
				&\;+ |c_1|^2\outprod{\alpha_2}_B\otimes \outprod{1}_M \otimes  D(\alpha_1\sqrt{1-\eta})\rho_G D^\dagger(\alpha_1\sqrt{1-\eta}) \\
				&\;+ c_0 c_1^*\lambda \proj{\alpha_0}{\alpha_1}_B\otimes \proj{0}{1}_M \otimes  D(\alpha_0\sqrt{1-\eta})\rho_G D^\dagger(\alpha_1\sqrt{1-\eta})\\
				&\;+ c_0^* c_1\lambda^* \proj{\alpha_2}{\alpha_1}_B\otimes \proj{1}{0}_M \otimes  D(\alpha_1\sqrt{1-\eta})\rho_G D^\dagger(\alpha_0\sqrt{1-\eta})
			\end{split}
		\end{align}
	\end{widetext}
	where $\rho_G=\outprod{\phi}_G$. Choosing $\alpha_0=-\alpha_1=\sqrt{\pi/1-\eta}/2$ in the ideal cavity interaction regime, the effective net displacement is either nullified (for memory state $\ket{0}_M$ ) or is of the magnitude $\sqrt{\pi}$. The former is an identity operation while the latter is a GKP basis Pauli $\hat{X}$ operation. Thus the entire action is an effective controlled $\hat{X}$ operation with the memory serving as the control qubit. Similarly, choosing $\alpha_0=-\alpha_1=\sqrt{\pi/1-\eta}/2$ results in an effective memory to GKP CPHASE gate. We subsequntly need to `unentangle' the bosonic mode from the joint system -- we perform a Pauli $\hat{X}$ gate on the memory and consider a secondary interaction of the bosonic mode with the coupled atom-cavity system. The bosonic mode components are transformed as 
	\begin{align}
		\begin{split}
			\ket{\alpha_0}_B\xrightarrow{r^{(1)}}\ket{r^{(1)*}(\omega)r^{(0)*}(\omega)\alpha f(\omega)}_B\\
			\ket{\alpha_1}_B\xrightarrow{r^{(0)}}\ket{r^{(0)*}(\omega)r^{(1)*}(\omega)\alpha f(\omega)}_B
		\end{split}
	\end{align}
	which nullifies any memory state dependent phase. The final joint memory-GKP state is then given by,
	{\small 
		\begin{align}
			\begin{split}
				\rho_{MG}\approx & \; |c_0|^2\outprod{1}_M \otimes  D(\alpha_0\sqrt{1-\eta})\rho_G D^\dagger(\alpha_0\sqrt{1-\eta})\\
				&\;+|c_1|^2 \outprod{0}_M \otimes  D(\alpha_1\sqrt{1-\eta})\rho_G D^\dagger(\alpha_1\sqrt{1-\eta}) \\
				&\; + c_0 c_1^* |\lambda|^2  \proj{1}{0}_M \otimes  D(\alpha_0\sqrt{1-\eta})\rho_G D^\dagger(\alpha_1\sqrt{1-\eta}) \\
				&\; + c_0^* c_1 |\lambda|^2  \proj{0}{1}_M \otimes  D(\alpha_1\sqrt{1-\eta})\rho_G D^\dagger(\alpha_0\sqrt{1-\eta})
			\end{split}
			\label{eq:gkp_spin_ent}
	\end{align}}
	Performing a final Pauli $\hat{X}$ gate on the memory yields the proper CNOT gate action; we shall represent the gate by $C_X^{M \rightarrow G}$.
	
	% In the ideal interaction limit where $\alpha_0\sqrt{1-\eta}=-\alpha_1\sqrt{1-\eta}=-\sqrt{\pi}/2$, with $|\lambda|^2=1$ the overall action is visualized by considering the initial states $\ket{\psi}_M= (\ket{0}_M\pm\ket{1}_M)/\sqrt{2}$ and $\ket{\phi}_G=\ket{\tilde{0}}_G$. The entire action generates the memory-GKP Bell pair 
	% \begin{align}
		% \begin{split}
			% 		\ket{\psi}_M\otimes \ket{\phi}_G \xRightarrow{C_X^{M \rightarrow G}} & \frac{\ket{0}_M\ket{\tilde{0}}_G\pm \ket{1}_M\ket{\tilde{1}}_G}{\sqrt{2}}\\
			% 	&\equiv \ket{\Phi^\pm}_{MG}
			% \end{split}
		% \end{align}
	For the required interaction to successfully apply the desired CNOT gate, we need to satisfy $\alpha_1\sqrt{1-\eta}\equiv \sqrt{\pi}/2$ and $ \alpha_0\sqrt{1-\eta}\equiv -\sqrt{\pi}/2$. Assuming $r^{(0)}(\omega)f(\omega )\rightarrow1$ and $r^{(1)}(\omega)f(\omega)\rightarrow-1 $, we need $\pm \alpha \sqrt{1-\eta}= \pm\sqrt{\pi}/2 \Rightarrow \alpha = \sqrt{\pi/(1-\eta)}/2$. The mean photon number content is $|\alpha|^2= {\pi}/({4(1-\eta)})$. To maintain the requirement that less than one photon interacts with the atom-cavity system over the interaction duration, we need the pulse length $ \tau$ to satisfy 
	\begin{align}
		\frac{\pi}{4(1-\eta)}\times\frac{1}{\kappa \tau} \ll 1
	\end{align}
	
	\begin{figure}
		\centering
		\includegraphics[width=\linewidth]{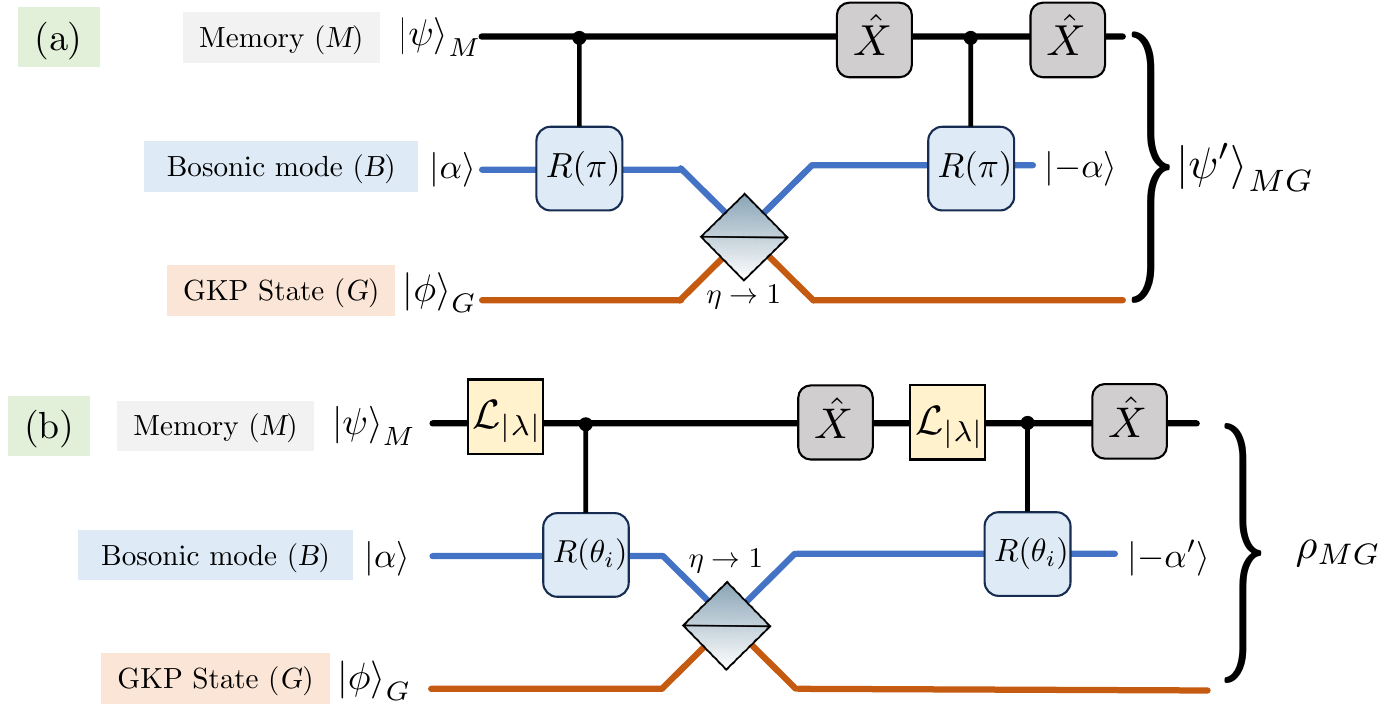}
		\caption{Effective circuit models for the (a) ideal interaction and (b) the realistic interaction. }
		\label{fig:circuit_model}
	\end{figure}
	
	We propose a simplified circuit model in Fig.~\ref{fig:circuit_model} to understand the effective $C_X^{M \rightarrow G}$ gate protocol. The memory dependent phase rotation gates are denoted by
	\begin{align}
		R(\theta_i)\equiv \exp\left[i \left(\outprod{0}_M\otimes \theta_0 \hat{a}^\dagger_B \hat{a}_B + \outprod{1}_M\otimes \theta_1 \hat{a}^\dagger_B \hat{a}_B\right)\right]
	\end{align}
	while the effective memory dephasing channel is $\mathcal{L}_{|\lambda|}$. The phase damping channel's action on an input state $\rho$ is given in the operator sum-representation by  $\mathcal{L}_\epsilon(\rho) =\sum_{i=0}^{1} K_i\rho K_i^\dagger$, where the Kraus operators are 
	\begin{align}
		K_0=\begin{bmatrix}
			1 & 0 \\
			0 & \epsilon
		\end{bmatrix}; \quad
		K_1= \begin{bmatrix}
			0 & 0 \\
			0 &\sqrt{1- \epsilon^2}
		\end{bmatrix}.
	\end{align}
	
	A close variant to the protocol presented above is the implementation of a effective CPHASE gate (which we may label $C_Z^{M\leftrightarrow G}$). Let us consider a coherent pulse $\ket{i\alpha}_B$ s.t. $ \alpha = \sqrt{\pi/(1-\eta)}/2$, and a pre-displaced GKP state $\ket{\phi'}_G=D(i\sqrt{\pi}/2)\left[\tilde{b}'_0 \ket{\tilde{0}}_G+ \tilde{b}'_1 \ket{\tilde{1}}_G\right] $ with  $\tilde{b}'_0,\tilde{b}'_1\in \mathbb{C}$ and $|\tilde{b}'_0|^2+|\tilde{b}'_1|^2=1$. The entire protocol sequence presented above now implements or nulls a conditional displacement of $\sqrt{\pi}/2$ along the $\hat{p}$ quadrature thus implementing a logical $C_Z$ operation. Consequently, both effective gates share operational features, requirements and limitations. The same is true for the hexagonal lattice GKP encoding~\cite{Noh2019-gd,Albert2018-qr}, where choosing $\alpha = \frac{1}{2} \sqrt{\frac{\pi}{\sqrt{3}}} \left( \frac{\sqrt{3}-i}{2}\right)$ with a pre-displacement of the initial state by $\frac{1}{2} \sqrt{\frac{\pi}{\sqrt{3}}} \left( \frac{\sqrt{3}-i}{2}\right)$ implements the effective $\effcnot$ gate.
	
	In an accompanying manuscript~\cite{Dhara2024-ru}, we show how entanglement can be generated between $N$-quantum memories and a GKP qudit of dimension $d=2^N$, by a generalized extension of this protocol.
	
	\section{Encoding Transduction between Memory and GKP Mode}
	\label{sec:bidirectional_transfer}
	
	In this section we explore circuits for qubit state transfer by leveraging the $\effcnot$ gate action. We characterize the final state outcome in terms of the realistic interaction model developed in the previous section. Furthermore we highlight the effect of measurement error induced degradation in final state quality for both teleportation circuits. 
	
	\begin{figure}
		\centering
		\includegraphics[width=0.9\linewidth]{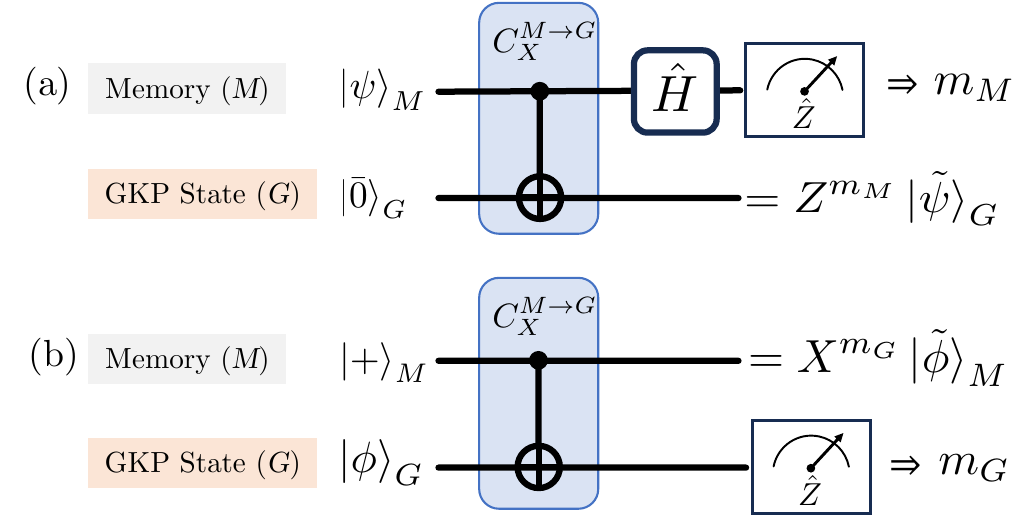}
		\caption{Effective single qubit circuits for state teleportation of quantum state from (a) memory to GKP mode, and (b) GKP to memory qubit. Each circuit outputs a transduced state (here shown to be pure) conditioned on the logical $Z$ measurement outcomes. }
		\label{fig:state_teleportation}
	\end{figure}
	
	\subsection{Memory to GKP State Transduction}
	\label{sec:transfer_mem2gkp}
	Starting with an initial memory state $\ket{\psi}_M=c_0 \ket{0}_M +c_1\ket{1}_M; c_0,c_1\in\mathbb{C} $ and an initial GKP state $\ket{{\phi}}_G=D(\sqrt{\pi}/2)\ket{\tilde{0}}_G$, the effective circuit is shown in Fig.~\ref{fig:state_teleportation} (a)~\cite{Nielsen2010-fr}. Upon reading out the memory in the Pauli $\hat{Z}$ basis the final state of the GKP qubit is
	\begin{align}
		\begin{split}
			\rho'_{G}=& |c_0|^2 D(\beta_0)\rho_G D^\dagger(\beta_0)  +|c_1|^2 D(\beta_1)\rho_G D^\dagger(\beta_1) \\
			& + (-1)^{m}|\lambda|^2( c_0 c_1^*  D(\beta_0)\rho_G D^\dagger(\beta_1) + c_0^* c_1  D(\beta_1)\rho_G D^\dagger(\beta_0))
		\end{split}
	\end{align}
	where $m=\{0,1\}$ is determined by the measurement outcome, $\beta_{\{0,1\}}=\alpha$ is the effective memory state dependent displacement on the GKP state and the other symbols have their corresponding meaning as defined in Sec.~\ref{sec:interaction_details}. The final state fidelity, after necessary Pauli frame correction, is given by evaluating $ \braket{\tilde{\psi}|\rho'_G|\tilde{\psi}}$ where $\ket{\tilde{\psi}}_G=c_0 \ket{\bar{0}}_G +c_1\ket{\bar{1}}_M$. 
	
	In realistic memory systems, measurements of the system can be error-prone due to non-idealities in the physical interface and operations. For e.g., in solid state qubits, measurements of the qubit is implemented by rotating the qubit basis and by a resonant readout pulse. The measurement outcome is deduced by recording the numbers of reflected photons or by phase-measurement; imperfections in the collection or detection can introduce a `readout error'. For the circuit under consideration here, given an intrinsic error probability $p_\mathrm{meas}$ for the memory measurement, the final state is $\tilde{\rho}_G=(1-p_\mathrm{meas})\rho'_G +p_\mathrm{meas}\, \hat{Z} \rho'_G \hat{Z}$.
	
	\subsection{GKP to Memory State Transduction}
	\label{sec:transfer_gkp2mem}
	Starting with an initial memory state $\ket{\psi}_M=(\ket{0}_M+\ket{1}_M)/\sqrt{2}$ and an initial GKP state $\ket{{\phi}}_G=D(\sqrt{\pi}/2)[\tilde{c}_0 \ket{\tilde{0}}_G +\tilde{c}_1 \ket{\tilde{1}}_G]$, the effective circuit is shown in Fig.~\ref{fig:state_teleportation} (b)~\cite{Nielsen2010-fr}. Upon reading out the GKP qubit in the Pauli $Z$ basis (i.e. via $q$ quadrature homodyne measurement) the final state of the memory qubit is
	\begin{align}
		\begin{split}
			\rho_{M}= &  |\tilde{c}_0|^2\outprod{0}_M +|\tilde{c}_1|^2 \outprod{1}_M  \\
			& \; + (-1)^{m} |\lambda|^2  \left[ \tilde{c}_0 \tilde{c}_1^*  \proj{0}{1}_M +  \tilde{c}_0^* \tilde{c}_1   \proj{1}{0}_M  \right]
		\end{split}
	\end{align}
	where $m_G=\{0,1\}$ is determined by the measurement outcome and the other symbols have their corresponding meaning as defined in Sec.~\ref{sec:interaction_details}. 
	We adopt a post-selected measurement strategy for the GKP qubit measurement as defined in Appendix~\ref{app:gkp_measurement}~\cite{Fukui2018-mz}.  We define $P_c$ and $P_f$ respectively, as tight lower bounds for the probability of a correctly interpreting the measurement outcome and for making a logical error~\cite{Rozpedek2023-vi}. These are defined by, 
	\begin{align}
		\begin{split}
			P_c&\coloneqq\int_{-\frac{1}{2} \sqrt{\pi}+v}^{\frac{1}{2} \sqrt{\pi}-v} f(x; 0,\sigma) d x; \\
			P_f&\coloneqq 2 \int_{\frac{1}{2} \sqrt{\pi}+v}^{\frac{3}{2} \sqrt{\pi}-v} f(x; 0,\sigma) d x,
		\end{split}
	\end{align}
	where $	f(x; \mu,\sigma )\coloneqq{\exp(-(x-\mu)^2/2\sigma^2)}/({\sigma \sqrt{2\pi}})$ is the standard Gaussian function. The state transfer from GKP to memory then succeeds with probability $P_c+P_f$, and the final state is given as $\tilde{\rho}_M= \left( P_c \, \rho_M +P_f \, \hat{Z} \rho_M \hat{Z} \right)/(P_c+P_f)$.
	
	\subsection{Teleportation Fidelity Analysis}
	\label{sec:teleport_fid}
	
	\begin{figure}[]
		\centering
		\includegraphics[width=\linewidth]{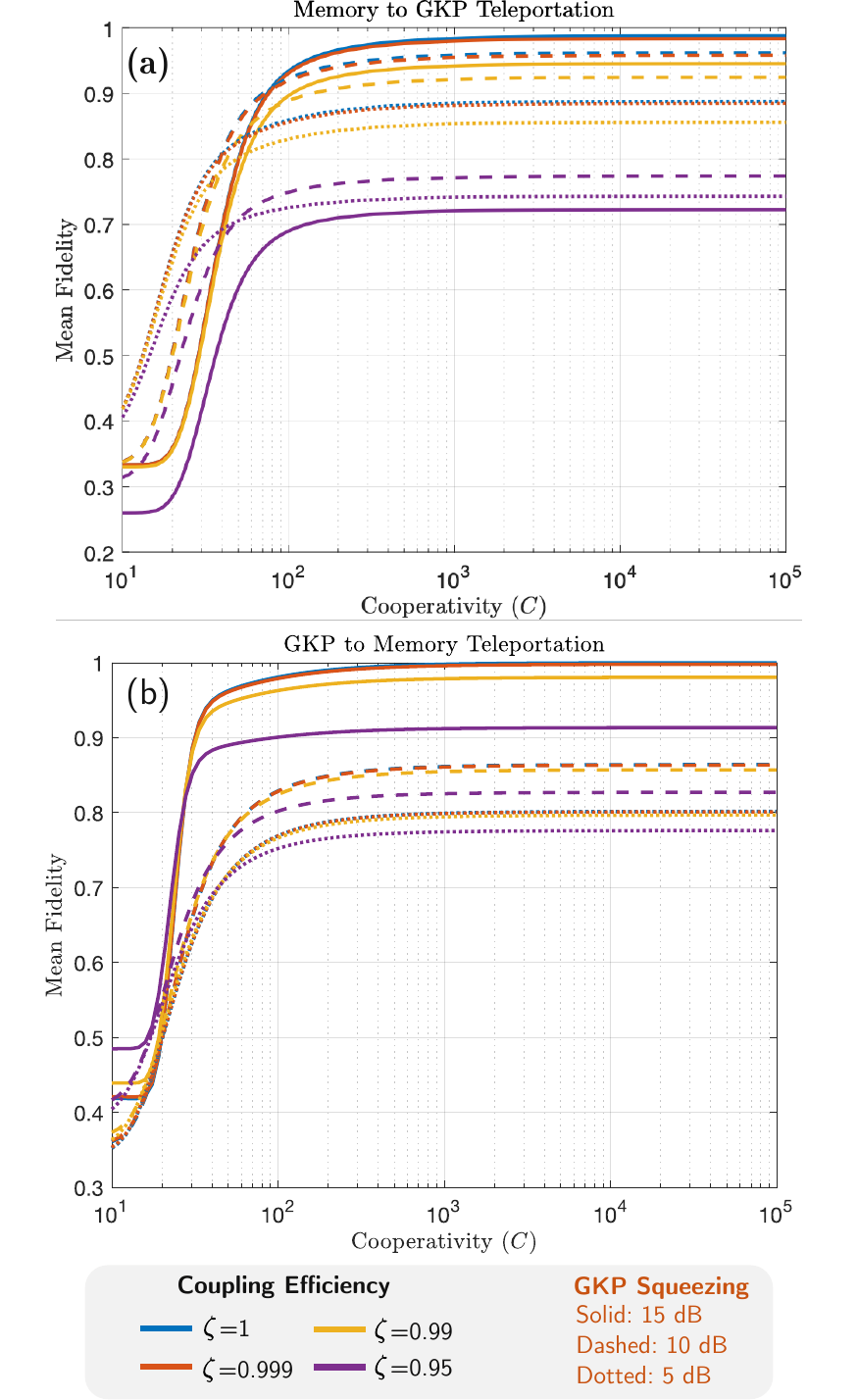}
		\caption{Mean fidelity of single qubit teleportation as a function of cavity cooperativity for the eigenstates of the Pauli $\hat{X},\hat{Y},\hat{Z}$ operators. We analyze bidirectional state teleportation from (a) memory to GKP qubit, and (b) GKP to memory qubit using the effective $\effcnot$ gate for varying values of cavity coupling efficiency ($\zeta$) and GKP qubit squeezing. The single qubit teleportation circuits are included in the inset of the plots.  }
		\label{fig:teleport}
	\end{figure}
	
	The fidelity of the transferred states is depicted in  Fig.~\ref{fig:teleport}, where we plot the average fidelity of teleportation (of the eigenstates of the Pauli $\hat{X},\hat{Y},\hat{Z}$ operators). The major source of infidelity in this protocol is the effective qubit dephasing ($|\lambda|^2$) as well as the lossy and imperfect phase rotation imparted by a sub-optimally coupled dispersive interaction. Fig.~\ref{fig:teleport} (a) and (b) respectively show the fidelity of teleporting the state of the memory qubit to the GKP basis and vice versa as function of $C$ for various values of $\zeta$ (assuming atomic detuning of $|\omega-\omega_a|=5\gamma_m$). We see that the overall mean teleportation fidelity is maximized for large values of $C\gtrsim100$, with the maximal value attained being inversely dependent on $\zeta$, which emphasizes the importance of the aforementioned strongly coupled low-loss cavity interaction requirement. Furthermore, as physical GKP qubit logical states are not truly orthogonal for low values of squeezing, their maximum attainable fidelity is lower than what is expected from their highly squeezed counterparts. For low values of cooperativity ($C\lesssim100$), the protocol implements an imperfect conditional displacement less than the required half-lattice spacing of $\sqrt{\pi}$. This results in the  low-squeezing GKP states (which have non orthogonal $\hat{Z}$ eigenstates due to their larger peak variances) having apparently better fidelity than their highly-squeezed counterparts (for the memory to GKP teleportation scenario) - this is not an optimal region of operation as the final qubit states are non-orthogonal.
	
	\section{GKP Cluster State Creation}
	\label{sec:cluster_generation}
	
	One possible application and a natural extension of the memory to GKP controlled NOT operation, is the creation of GKP cluster states. One can achieve this in multiple ways. We present two variants that we shall call the 'sequential interaction' (\texttt{seq-int}) and the `recurrent interaction' (\texttt{rec-int}) protocols.
	
	In the \texttt{seq-int} protocol, we entangle the GKP states to one another by the $\effcnot$ action via a train of temporally subsequent coherent pulses. The input mode carrying the GKP qubits must also be excited in this temporal sequence with the corresponding GKP qubits required to create a massive entangled state of GKP qubits with the memory. For a target cluster state, each edge on the state representation graph corresponds to an additional interaction.  Any natural extension of the effective circuits in Fig.~\ref{fig:circuit_model} can be implemented in principle -- additional inter-pulse memory qubit rotations must be accounted for wherever necessary. Subsequent measurement of the memory qubit in the Pauli $X$ basis will disentangle the memory from the cluster state of GKP qubits. The \texttt{seq-int} protocol is particularly useful for generating linear cluster states - to introduce any `branching' of the state, additional interactions are necessary. Readers must note that for each realistic $ \effcnot$ interaction, the system imparts a qubit phase decay of magnitude $|\lambda|^2$ -- thus for a $n$-GKP qubit cluster, the phase decay will be $\propto |\lambda|^{2n}$. Additionally the time taken for the  \texttt{seq-int} protocol scales linearly with the cluster state size -- given some temporal bandwidth $\tau$ for each coherent pulse, the single $\effcnot$ gate lasts atleast $3\tau$ seconds (two interactions with the cavity and a single beamsplitter interaction). Thus the creation of a $n$-connection GKP qubit cluster will take $3n\tau$ seconds.
	
	The \texttt{rec-int} protocol uses the bosonic state (entangled with the memory; Eq.~\eqref{eq:spin_coh_st}) in a repeated fashion for displacements of the different GKP states. The sequential displacement can be implemented via multiple consecutive nearly reflective beamsplitters where the complementary modal inputs are the different GKP states required to form a cluster. Alternatively by the use of re-configurable optical paths, the bosonic mode can be re-routed through the same beamsplitter with an intermediate temporal delay (i.e.\ an optical path length difference given by the temporal separation of subsequent GKP qubits). The effective qubit dephasing induced via this \texttt{rec-int} protocol is (in-principle) agnostic to the cluster size - with ideal displacement unitary action, the net dephasing factor on the memory qubit is only $|\lambda|^2$ since only pair of interactions are required. This protocol is well suited to the creation of a highly-connected cluster state (for e.g. an all-to-all connected clique graph). Furthermore, the repetition time in this case will also scale less severely with the size of the cluster $n$ as $(n+2)\tau$. 
	
	Readers must note that throughout the previous section, as well as the discussion above, we assume ideal displacement operations on the GKP qubits by the beamsplitter interactions. In a realistic scenario, the reflectivity $\eta$ of the beamsplitter can never be arbitrarily close to 1 and is practically limited to some $\eta=1-\epsilon$. The minimum achievable $\epsilon$ is limited by realistic beamsplitter unitary implementations, and more importantly by temporal bandwidth limitations on the coherent pulse. Since $\tau\gg\pi/(4(1-\eta)\kappa)=\pi/(4\epsilon\kappa)$, we are limited by both the coherence time of the input pulse (i.e.\ the laser coherence time if the coherent pulse is optical) as well as any qubit decoherence mechanism in the atomic state (which we have not considered in this article). Furthermore, the coherent state $\ket{\alpha}_B$ is more accurately modelled by a thermalized state $\ket{\alpha;\bar{n}}_B= D(\alpha)\rho_{\mathrm{th}}(\bar{n})D^\dagger(\alpha)$ i.e. a displaced thermal state $\rho_{\mathrm{th}}(\bar{n}) = \sum_{k=0}^{\infty} \, [\bar{n}^k/(\bar{n}+1)^{k+1}] \outprod{k}$ of mean photon number per mode of $\bar{n}$. We get to the ideal coherent state for $\bar{n}=0$; however for $\bar{n}>0$ the effective displacement circuit also introduces an additional `thermalization' of the mode being mixed (see Appendix~\ref{app:effective_disp_interaction} for detailed explanations). For non-infinitesimal values of beamsplitter transmissivity $\epsilon$, the target state also undergoes pure loss and the coherent state output mode will be affected by complementary input mode (in this case carrying the pre-displaced GKP states) due to finite leakage of the state. This acts effectively as further thermal noise on the bosonic mode. Hence, for the \texttt{rec-int} protocol, the performance of each interaction degrades over time, thereby imposing a limit on the maximum cluster size. The \texttt{seq-int} protocol method is less prone to this; however a detailed study of the effect of GKP qubit leakage into the coherent state is beyond the scope of this study and left for future work.
	
	\section{GKP Qubit Assisted Entanglement Generation Between Memories}
	\label{sec:gkp_assisted_swap}
	\subsection{Ideal GKP Qubit Assisted Swap}
	Entanglement generation between remote quantum memories is of great importance to many applications in quantum information processing. One possible way to generate entanglement is to entangle bosonic qubits with the memories followed by a Bell projection on the bosonic qubits to entangle the memories~\cite{Barrett2005-sm,Dhara2023-zv}. Swaps adopting discrete variable encodings of qubits and linear bosonic interactions (i.e. balanced two-mode mixing or beamsplitters) with photon number resolved detections are fundamentally probabilistic~\cite{Barrett2005-sm,Calsamiglia2001-vp,Dhara2023-zv} - with ideal hardware and no ancilla resources (for e.g. single photons or inline squeezing) they succeed with probability $1/2$ with successful swaps being heralded by the measurement outcomes. Here we propose the use of GKP qubits entangled to the memories for this mediated swap. Unlike discrete-variable encodings, the Bell projection in the GKP qubit basis comprises of  (1) two-mode interaction on a balanced beamsplitter, followed by (2) dual-homodyne measurement of the two modes along orthogonal quadratures~\cite{Walshe2020-va,Weedbrook2012-xk}. Since homodyne measurements always yield an usable outcome, the swap is inherently deterministic. However one may relax the determinism of the swapping process to yield better quality entangled states at the cost of a reduced probability of success~\cite{Fukui2018-mz,Fukui2021-ay}.
	
	\begin{figure}
		\centering
		\includegraphics[width=0.7\linewidth]{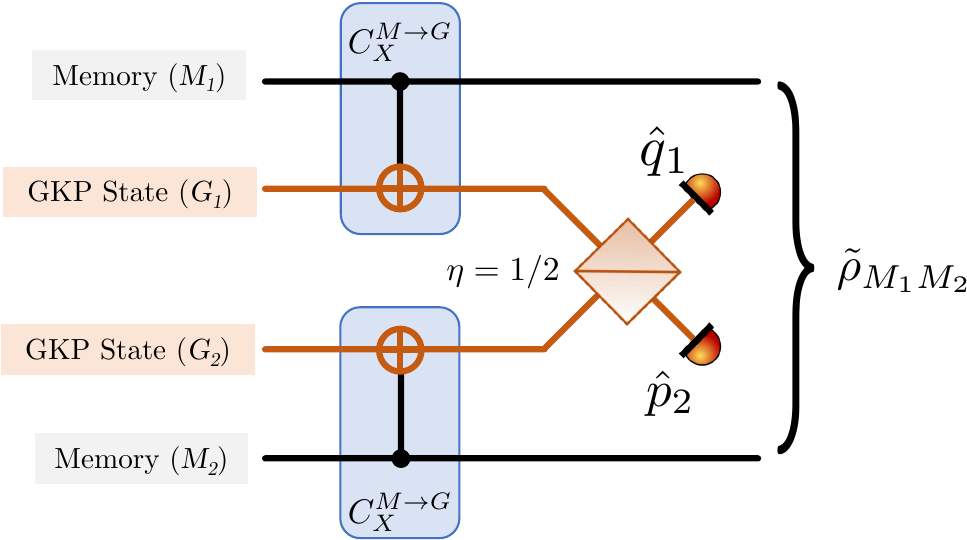}
		\caption{Circuit for the generation of spin-spin entanglement using GKP assisted swaps. }
		\label{fig:ent_swap_model}
	\end{figure}
	
	Given two input GKP states $\ket{\phi_1}_G$ and $\ket{\phi_2}_G$, the action of the balanced beamsplitter (which we denote by $\mathcal{B}_{1,2}$) followed by the dual-homodyne measurement (with two outcomes say $m_1$ and $n_2$), the following identity can be shown to be true~\cite{Walshe2020-va,Fukui2021-ay} -- 
	\begin{align}
		\begin{split}
			\bra{m_1}_q\bra{n_2}_p \mathcal{B}_{1,2} \ket{  \phi_1,\phi_2}_G = \bra{n_2}_p\bra{m_1}_q C_{X}^{1\rightarrow2} \ket{  \phi_1,\phi_2}_G,
		\end{split}
	\end{align}
	where $C_X^{1\rightarrow2}=\exp(-i\hat{q}_1\hat{p}_2)$ is the GKP basis $\mathrm{CNOT}$ gate. 
	
	We examine this through an example -- consider a scenario where the initial spin state is $\ket{\psi}_m=(\ket{0}_M+\ket{1}_M)/\sqrt{2}$ and the GKP state is initialized as $\ket{\phi}_G=D(\sqrt{\pi}/2) \ket{\tilde{0}}_G$. Assuming we satisfy the requirements for an ideal memory to GKP qubit $C_X^{M\rightarrow G}$ gate, the final joint state of the memory-GKP qubits is $(\ket{0}_M\ket{\tilde{0}}_G+\ket{1}_M \ket{\tilde{1}}_G)/\sqrt{2}\equiv\ket{\Phi^+}_{MG}$. Considering two copies $\ket{\Phi^+}_{MG}$, the effective GKP basis $C_X^{1\rightarrow2}$ gate action is
	\begin{align}
		\begin{split}
			&\ket{\Phi^+}_{M_1 G_1}\otimes \ket{\Phi^+}_{M_2 G_2}\\
			& \quad \xRightarrow{C_X^{1\rightarrow2}} \ket{\Phi^+}_{M_1M_2}\ket{\tilde{+},\tilde{0}}_{G_1G_2} +  \ket{\Psi^+}_{M_1M_2}\ket{\tilde{+},\tilde{1}}_{G_1G_2},
		\end{split}
	\end{align}
	Subsequently the dual homodyne measurement of the modes along $\hat{p}_1$ and $\hat{q}_2$ quadratures will yield random real-valued outcomes $x_1$ and $y_2$~\cite{Gottesman2001-hd,Fukui2018-mz,Rozpedek2023-vi}. Standard GKP measurement proceeds by processing these outcomes in terms of their integral peak position i.e. the value of $\lfloor x/\sqrt{\pi} \rfloor$ where $x\in\{p_1,q_2\}$, informs us of the logical outcomes. Since $\hat{p},\hat{q}$ homodyne measurements of the modes are equivalent to performing GKP basis Pauli $X_1 $ and $Z_2$ measurements on the two qubit states. For the state derived above, we note that there are two possible outcomes $\{+,0\}$ or $\{+,1\}$, which project the states of the memories into $\ket{\Phi^+}_{M_1 M_2}$ or $\ket{\Psi^+}_{M_1 M_2}$.
	
	\subsection{Physical GKP Qubit Assisted Swaps}
	An important caveat for the homodyne measurement is that it has a non-zero probability of resulting in a logical error; this arises from the finite Gaussian peak width in physical GKP qubits~\cite{Gottesman2001-hd,Fukui2018-mz,Rozpedek2023-vi}. Given a finitely squeezed GKP state parameterized by the state squeezing $ \sigma_G^2$, the measurement error probability varies inversely with $ \sigma_G^2$. Post-selection of the measurement outcomes~\cite{Fukui2018-mz} is one way to boost the final state quality at the cost of a sub-unity measurement success probability. This was briefly introduced in Sec.~\ref{sec:transfer_gkp2mem} and is covered in detail in Appendix~\ref{app:gkp_measurement}. If the final desired state is $\rho_{M_1 M_2} $, the final state $	\tilde{\rho}_{M_1 M_2}$ after post-selected dual-homodyne measurements is heralded with a success probability $(P_c+P_f)^2$ and is described by,
	\begin{align}
		\tilde{\rho}_{M_1 M_2}=\frac{P_c^2 \, \rho_{M_1M_2}+  P_c P_f \, \rho^{(1)}_{M_1M_2} + P_f^2 \rho^{(2)}_{M_1M_2}}{(P_c+P_f)^2}, 
		\label{eq:spinspin_ent2}
	\end{align}
	where $\rho^{(1)}_{M_1 M_2}=\mathbb{I}_1 X_2\, \rho_{M_1 M_2}\, \mathbb{I}_1 X_2 + Z_1 \mathbb{I}_2\,  \rho_{M_1 M_2} \, Z_1 \mathbb{I}_2 $ and $\rho^{(2)}_{M_1 M_2}=Z_1X_2 \, \rho_{M_1 M_2} Z_1 X_2 $. As defined $P_c$ is the lower bound to probability of a correctly interpreting the measurement outcome, and $P_f$ is lower bound to the probability of a logical error in the measurement. The fidelity of this state to the desired state is given by
	\begin{align}
		F(\tilde{\rho}_{M_1 M_2}, \rho_{M_1 M_2}) = P_c^2/(P_c+P_f)^2.
	\end{align}
	
	When $\rho_{M_1,M_2}$ is one of the Bell pairs $\{\ket{\Phi^\pm}_{M_1,M_2}, \ket{\Psi^\pm}_{M_1,M_2}\}$, we may use the hashing bound~\cite{Devetak2005-hi} as an achievable lower bound to distillable entanglement of the state (i.e. number of Bell pairs that can distilled from a copy of state given both parties possess ideal quantum computers). The hashing bound is defined as  $I(\rho)=\max[S(\rho_A)-S(\rho_{AB}),S(\rho_B) - S(\rho_{AB})]$ where $S(\rho)$ is the von Neumann entropy of the state $S(\rho)=-\Tr (\rho \log_2\rho )$. The hashing bound of the final state is then given by,
	\begin{align}
		I(\tilde{\rho}_{M_1 M_2}) = 1+ \sum_{i=1}^4 p_i \log_2 p_i,
	\end{align}
	where 
	$p_1=  P_c^2/(P_c+P_f)^2; \; p_2=p_3=  P_c P_f/(P_c+P_f)^2; \; p_4= P_f^2/(P_c+P_f)^2$.
	
	\begin{figure}
		\centering
		\includegraphics[width=\linewidth]{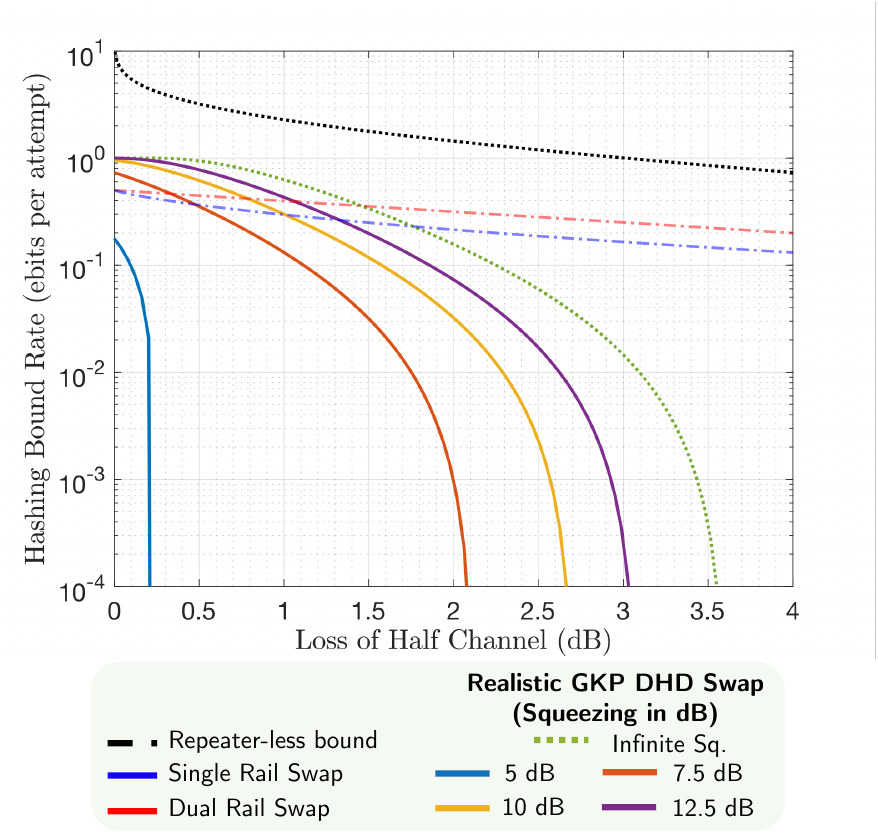}
		\caption{Comparison of the hashing bound rate as function of the half channel loss $(\sqrt{\eta})$ of GKP-assisted entanglement generation between quantum memories with the repeaterless bound (black; dash-dot) and the best achievable rate  with single rail (blue; dash-dot) and dual rail (red; dash-dot) encoding assisted swaps. Performance of physical GKP qubits (colors) is shown along with a rate envelope (green; dashed) which shows the best achievable performance with infinitely-squeezed GKP qubits $(\sigma^2_\mathrm{GKP}\rightarrow 0)$. }
		\label{fig:gkp_swap}
	\end{figure}
	
	We compare the performance of GKP-qubit assisted entanglement generation links in the `midpoint swap' configuration to links employing DV encoded qubits and the ultimate repeaterless bound in  Fig.~\ref{fig:gkp_swap}. Here, we assume that all detectors and interference circuits are ideal and noiseless, i.e., we only account for the effect of loss for the GKP qubits. We plot the hashing bound rate (units: ebits per attempt) versus the half channel loss $(\sqrt{\eta})$ for the various swaps. The repeater-less bound (black; dash-dot) to channel capacity is the ultimate upper bound for the entanglement generation rate (in ebits per channel use) between two parties communicating over a pure loss channel --- it is given by $D_2(\eta)=-\log_2(1-\sqrt{\eta})$. The best achievable rate of the single rail (blue; dash-dot) and dual rail (red; dash-dot) encoding assisted swaps are shown for comparison~\cite{Dhara2023-zv}. Performance of physical GKP qubits (varied colors) with a fixed post-selection window (varied line styles) is shown along with a rate envelope (green; dashed) which shows the best achievable performance with infinite-squeezed GKP qubits $(\sigma^2_\mathrm{GKP}\rightarrow 0)$ and a maximum post-selection window $(1-v_{\max}=10^{-4}) $. 
	
	Readers should note that both DV encodings are limited to a maximum rate of 0.5 (at $\eta=0$ dB) ebit per attempt, whereas the best performance achievable by (highly squeezed) physical GKP qubits is 1 ebit per attempt, demonstrating an inherent benefit for the proposed approach. GKP-assisted links surpass the performance of the analyzed DV encodings upto $\sim$3 dB loss - beyond that we note a sharp roll-off the in performance of GKP assisted swaps whereas the DV-qubit assisted links can sustain a positive entanglement generation rate. The roll-off in the rate for GKP-assisted links is quite similar to the effect of excess photons in the channel (for e.g. as shown in Ref.~\cite{Dhara2023-zv}) --- this is primarily due to the fact that the effect of photon loss in GKP qubits is similar to introducing excess noise on the detectors~\footnote{Loss causes the GKP peak variance to expand and causes an overall contraction of the grid-like state in phase space. The loss of `peak resolution' will induce measurement errors.}. However, post-selection of the measurement outcomes by choosing a larger rejection window $v\rightarrow1$ allows the link to maintain positive non-zero rates. In Fig.~\ref{fig:gkp_swap}, the green dotted curve denotes this rate envelope achieved by ideal GKP qubits for a maximum post-selection window of $v_{\max}=1-10^{-4}$. By choosing an optimal $v_{\max}$ which is closer to 1, we can extend the range of envelope indefinitely.
	
	In an accompanying manuscript~\cite{Dhara2024-ru}, we show how the maximally entangled states of $N$ quantum memories and GKP qudits ($d=2^N$) along with dual-homodyne entanglement swaps that retain analog information, enables quantum links that can sustain channel capacity-approaching rates at low loss.
	% {
		% \color{blue}
		% Plots to be included
		% - Ultimate rate of swap compared with single rail and dual rail swaps
		% include effect of finite squeezing and practical post-selection windows
		% - 
		% }

	\section{Conclusions and Outlook}
	\label{sec:conclusions}
	
	In this article we present a protocol for the implementation of an hybrid entangling operation between a quantum memory and a GKP qubit. The detailed analysis of the interaction and the operational requirements in terms of the sub-system parameters is developed. Our protocol enables a scalable approach for the interface of various qubit encodings. We list a few potential applications based on the analyzed use-cases that might be of interest to the community --
	\begin{enumerate}
		\item Quantum State Transfer (Encoding Transduction)
		\begin{itemize}
			\item GKP qubit decoding and error correction.
			\item Entanglement between CV and dual-rail encoded qubits.
		\end{itemize}
		\item GKP Cluster State Generation
		\begin{itemize}
			\item Raussendorf-Harrington-Goyal lattice creation for fault tolerant measurement-based quantum computing.
			\item Stabilizer code creation for error corrected quantum repeaters.
		\end{itemize}
		\item GKP Swap Assisted Memory Entanglement
		\begin{itemize}
			\item Memory qubit cluster state generation via percolation.
			\item Stabilizer code assisted entanglement overhead reduction.
		\end{itemize}
		\item CV Qubit Breeding Protocols
		\begin{itemize}
			\item Resource optimized generation of optimal CV qubits/qudits.
			\item Memory assisted qubit state correction for state generated by Gaussian Boson sampling.
		\end{itemize}
	\end{enumerate}
	
	We hope that our protocol will provide illuminating guidelines for experimental demonstrations and further research into advanced potential applications. The extension of our protocol for an effective interface between higher dimensional CV encodings and a memory system is left for future studies.
	
	\section*{Acknowledgments}
	The authors would like to thank Filip Rozpedek (UMass.\ Amherst), Kaushik P.\ Seshadreesan (University of Pittsburgh), Paul Polakos (Cisco), Nicolas Cerf (Universit\'{e} Libre de Bruxelles), Kevin C. Chen (MIT; currently at HRL Laboratories), and Hyeongrak Choi (MIT; currently at Stony Brook University) for fruitful discussions and comments on the manuscript. P.D. and S.G. acknowledge the Mega Qubit Router (MQR) project funded under federal support via a subcontract from the University of Arizona Applied Research Corporation (UA-ARC), for supporting this research. Additionally, all authors acknowledge the Engineering Research Center for Quantum Networks (CQN), awarded by the NSF and DoE under cooperative agreement number 1941583, for synergistic research support. L.J. additionally acknowledges support from the AFOSR MURI (FA9550-19-1-0399, FA9550-21-1-0209, FA9550-23-1-0338), DARPA (HR0011-24-9-0359, HR0011-24-9-0361), NSF (OMA-1936118, OMA-2137642, OSI-2326767, CCF-2312755), NTT Research, Packard Foundation (2020-71479), and the Marshall and Arlene Bennett Family Research Program.
	
%	S.G. has outside interests in SensorQ Technologies Incorporated and Guha, LLC. These interests have been disclosed to UA  and reviewed in accordance with their conflict of interest policies, with any conflicts of interest to be managed accordingly.
	
	\onecolumngrid
	
	\renewcommand\thefigure{\thesection.\arabic{figure}}  
	\counterwithin{figure}{section}  
	\appendix 
	
	\section{Langevin Equations: Derivation and Simplification}
\label{app:cavity_atom_langevin}
We consider the interaction of photons in bosonic modes with atom coupled to a single sided cavity. All parameters of the systems under consideration  are summarized below
\begin{itemize}
	\item Atomic System Parameters
	\begin{itemize}
		\item Energy eigenlevels: $ \ket{0}_M,\ket{1}_M $ and $\ket{2}_M  $.
		\item Eigenenergies: $ \hbar\omega_0,\hbar\omega_1,\hbar\omega_2 $.
		\item Coupled transition: $ \ket{1}_M\leftrightarrow\ket{2}_M $
		\item Atomic decay $\ket{2}_M\!\rightarrow\!\ket{1}_M  $: Decay occurs into (un-detected) mode  $\hat{a}_\gamma$ with rate $\gamma_m$. 
		\item \emph{All other decay mechanisms are forbidden or have a negligible rate.}
	\end{itemize}
	\item Cavity Modal Parameters
	\begin{itemize}
		\item Incoming, outgoing and cavity modes : $ \hat{a}_\mathrm{in}, \hat{a}_\mathrm{out} $ and $\hat{a}_c$. 
		\item Cavity input-output relation $\hat{a}_\mathrm{out}=\sqrt{\kappa_c}\hat{a}_c+\hat{a}_\mathrm{in}$.
		\item Atom-cavity coupling: $\hbar g(\hat{a}_c \hat{\sigma}_{21} + \hat{a}^\dagger_c\hat{\sigma}_{12}).$
		\item Cavity loss: Lost photons are coupled into an auxiliary (un-detected) mode $ \hat{a}_l $ with a coupling rate $ \kappa_l $.
	\end{itemize}
\end{itemize}

With the above considerations the cavity, atom and interaction Hamiltonians are given as
\begin{align}
	\begin{split}
		H_{\mathrm{cav.}}&=\hbar \omega_c \hat{a}^\dagger_c\hat{a}_c;\\
		H_{\mathrm{atom}}&= \sum_{i=0}^{2} \hbar \omega_i \outprod{i}_M;\\
		H_{\mathrm{int.}}&=\hbar g(\hat{a}_c \hat{\sigma}_{21} + \hat{a}^\dagger_c\hat{\sigma}_{12}).
	\end{split}
\end{align}
where $ \omega_c $ is the cavity resonance frequency, transition frequency and $ g $ is the cavity mode-atom coupling strength. The complete Hamiltonian is then 
\begin{align}
	\hat{H}={H_{\mathrm{cav.}}}+{H_{\mathrm{atom}}}+{H_{\mathrm{int.}}}
\end{align}

We analyze the system using the Heisenberg-Langevin formulation for the cavity and atomic operators~\cite{Gardiner1985-hn} to derive effective mode transformation relations. 
For the cavity annihilation operator $\hat{a}_c$, the bath operators are $\hat{a}_\mathrm{out}$ and $\hat{a}_l $, yielding the Langevin equation
\begin{align}
	\begin{split}
		\dot{\hat{a}}_c=\frac{-i}{\hbar} \left[\hat{a}_c,\hat{H}\right]-\sqrt{\kappa_c}\hat{a}_\mathrm{out}-\sqrt{\kappa_l}\hat{a}_l+\frac{\kappa}{2}\hat{a}_c.
		\label{eq:lgvn_cavity}
	\end{split}
\end{align}
For the atomic coherence operator $\hat{ \sigma}_{12} $ the bath operator is $\hat{a}_\gamma$; yielding the Langevin equation
\begin{align}
	\begin{split}
		\dot{\hat{ \sigma}}_{12}&=-\frac{i}{\hbar} [\hat{\sigma}_{12},\hat{H}]-\left[ \left[\hat{ \sigma}_{12}, \hat{ \sigma}_{21}\right] \left(-\frac{\gamma}{2} \hat{\sigma}_{12}+\sqrt{\gamma} \anh[\gamma]{a}(t) \right) -\left(-\frac{\gamma}{2} \hat{\sigma}_{21}+\sqrt{\gamma} \crt[\gamma]{a}(t)  \right)[\hat{ \sigma}_{12}, \hat{ \sigma}_{12}] \right]\\
		&=-\frac{i}{\hbar} [\hat{\sigma}_{12},\hat{H}]+(\hat{\sigma}_{22}-\hat{\sigma}_{11}) \left(-\frac{\gamma}{2} \hat{\sigma}_{12}+\sqrt{\gamma} \anh[\gamma]{a}(t)  \right).
		\label{eq:lgvn_atomic}
	\end{split}
\end{align}
Substituting the complete Hamiltonian $\hat{H}$ into these equations to simplify the relations, we obtain,
\begin{subequations}
	\begin{align}
		\dot{\hat{a}}_c &= -i\omega_c\hat{a}_c-ig\hat{\sigma}_--\sqrt{\kappa_c}\hat{a}_\mathrm{out}-\sqrt{\kappa_{l}}\hat{a}_l+\frac{\kappa}{2}\hat{a}_c;\\
		\dot{\hat{\sigma}}_- &=-i\omega_a\hat{\sigma}_- + ig\hat{a}_c\hat{\sigma}_z-\frac{\gamma}{2}\hat{\sigma}_z \hat{\sigma}_-+ \sqrt{\gamma} \hat{a}_\gamma\hat{\sigma}_z.
	\end{align}
\end{subequations}
Taking the operator Fourier transform, we get
\begin{subequations}
	\begin{align}
		\begin{split}
			& -i\omega {\hat{a}}_c (\omega) = -i\omega_c\hat{a}_c(\omega)-ig\hat{\sigma}_- -\sqrt{\kappa_c}\hat{a}_\mathrm{out}(\omega)-\sqrt{\kappa_{l}}\hat{a}_l(\omega)+\frac{\kappa}{2}\hat{a}_c(\omega)\\
			& \Rightarrow  \left(i \Delta_c-\frac{\kappa}{2}\right)\hat{a}_c(\omega) = -ig\hat{\sigma}_- -\sqrt{\kappa_c}\hat{a}_\mathrm{out}(\omega)-\sqrt{\kappa_{l}}\hat{a}_l(\omega); \label{eq:freq_dom1}
		\end{split}\\
		\begin{split}
			&	-i\omega {\hat{\sigma}}_- =-i\omega_a\hat{\sigma}_- + ig\hat{a}_c (\omega) \hat{\sigma}_z-\frac{\gamma}{2}\hat{\sigma}_z \hat{\sigma}_- + \sqrt{\gamma} \hat{a}_\gamma (\omega) \hat{\sigma}_z \\
			&\Rightarrow	\left(i\Delta_a +\frac{\gamma}{2}\hat{\sigma}_z\right) \hat{\sigma}_- =  ig \hat{a}_c(\omega)\hat{\sigma}_z + \sqrt{\gamma} \hat{a}_\gamma\hat{\sigma}_z .\label{eq:freq_dom2}
		\end{split}
	\end{align}
\end{subequations}
% With the definitions $ \Delta_c=\omega_c-\omega;\; \Delta_a=\omega_a-\omega $, upon rearranging we get,
% \begin{subequations}
	% 	\begin{align}
		% 		\\
		% 		\left(i\Delta_a +\frac{\gamma}{2}\hat{\sigma}_z\right) \hat{\sigma}_- &=  ig \hat{a}_c(\omega)\hat{\sigma}_z + \sqrt{\gamma} \hat{a}_\gamma\hat{\sigma}_z \label{eq:freq_dom2}
		% 	\end{align}
	% \end{subequations}
Assuming $ \langle\sigma_z\rangle= -1 $ (i.e. atom is almost always in the ground state for sufficiently large $\Delta_a$), we may simplify Eq.~\eqref{eq:freq_dom2} as
\begin{align}
	\hat{\sigma}_-=\frac{ig \hat{a}_c(\omega)+\sqrt{\gamma}\hat{a}_{\gamma}}{\gamma/2-i\Delta_a},
\end{align}
which when used in Eq.~\eqref{eq:freq_dom1} yields,
\begin{align}
	\hat{a}_{c}(\omega)=\frac{\sqrt{ \kappa_{c}} \hat{a}_{\text {out }}(\omega)+\sqrt{\kappa_{l}} \hat{a}_{l}(\omega)+ \frac{i g \sqrt{ \gamma}}{\gamma/2-i \Delta_{a}} \hat{a}_{\gamma}(\omega)}{\kappa/2-i \Delta_{c}+\frac{g^{2}}{\gamma/2-i \Delta_{a}}}.
\end{align}
This gives us the following equation when substituted in the input-output relation 
\begin{align}
	\begin{split}
		&\hat{a}_\mathrm{out}-\hat{a}_\mathrm{in}=\sqrt{\kappa_c}\times \frac{\sqrt{ \kappa_{c}} \hat{a}_{\text {out }}(\omega)+\sqrt{ \kappa_{l}} \hat{a}_{l}(\omega)+ \frac{i g \sqrt{ \gamma}}{\gamma/2-i \Delta_{a}} \hat{a}_{\gamma}(\omega)}{\kappa/2-i \Delta_{c}+\frac{g^{2}}{\gamma/2-i \Delta_{a}}}\\
		&\Rightarrow
		\left({\kappa/2-i \Delta_{c}+\frac{g^{2}}{\gamma/2-i \Delta_{a}}} \right)(\hat{a}_\mathrm{out}-\hat{a}_\mathrm{in})= \kappa_c \hat{a}_\mathrm{out}+\sqrt{\kappa_c\kappa_l}\hat{a}_l(\omega) + \frac{i g \sqrt{ \gamma \kappa_{c}}}{\gamma/2-i \Delta_{a}} \hat{a}_{\gamma}(\omega)
	\end{split}
\end{align}
Solving for $ \hat{a}_\mathrm{in} $, we obtain the overall input-output relation 
\begin{subequations}
	\begin{align}
		\hat{a}_{\text {in }}(\omega)&=\frac{ \left(\kappa/2-i \Delta_{c}+\frac{g^{2}}{\gamma/2-i \Delta_{a}}- \kappa_{c}\right) \hat{a}_{\text {out }}(\omega)-\sqrt{\kappa_{l} \kappa_{c}} \hat{a}_{l}(\omega)-\frac{i g \sqrt{\gamma \kappa_{c}}}{\gamma/2-i \Delta_{a}} \hat{a}_{\gamma}(\omega)}{\kappa/2-i \Delta_{c}+\frac{g^{2}}{\gamma/2-i \Delta_{a}}}\\
		&=r(\omega) \cdot \hat{a}_{\text {out }}(\omega)+l_C(\omega) \cdot \hat{a}_{l}(\omega)+ l_A(\omega) \cdot \hat{a}_{\gamma}(\omega)
	\end{align}
\end{subequations}
Considering cavity resonant pulses $\Delta_c=0$, we obtain
\begin{subequations}
	\begin{align}
		r(\omega) &=1-\frac{2 \zeta}{C'+1}; \\
		l_C(\omega) &=-\frac{2 \sqrt{\zeta(1-\zeta)}}{C'+1}; \\
		l_A(\omega) &=\frac{-2iC' \sqrt{2 \zeta/ C}}{C'+1},
	\end{align}
\end{subequations}
where, we substitute $ 4g^2/\kappa \gamma \equiv C$ for the cavity cooperativity, and use the cavity coupling efficiency $ \zeta=\kappa_c/\kappa $ along with the modified cooperativity $C'=C/(1-2i\Delta_a/\gamma_m) $. Readers may note that the virtual excitation condition (i.e.\ $ \langle\sigma_z\rangle= -1 $) is not true~\cite{Raymer2024-nz} when $\Delta_a\rightarrow0$; however the cavity-atom system still imparts a memory dependent phase on the travelling modes~\cite{Duan2004-qs}.
	
	\section{Gottesman-Kitaev-Preskill States: Definition and Post-Selected Homodyne Measurements}
\subsection{Definitions}
\label{app:gkp_definition}
We consider the physical GKP qubits which have finite mean photon number (i.e. finite energy)  and are properly normalized. In the position and momentum representation the states are defined as 
\begin{subequations}
	\begin{align}
		\ket{\tilde{L}}_{G}\propto & \sum_{t=-\infty}^{\infty}\int e^{-\pi\sigma_2^2(2t+L)^2/2} \exp\left[-\frac{(q-(2t+L)\sqrt{\pi})^2}{2\sigma_1^2}\right]\ket{q} dq\\
		\propto &\sum_{t'=-\infty}^{\infty}\int (\kappa_L)^{t'} e^{-\pi\sigma_1^2t'^2/2} \exp\left[-\frac{(p-t'\sqrt{\pi})^2}{2\sigma_2^2}\right]\ket{p} dp; \; 
	\end{align}
\end{subequations}
where $\kappa_L=(-1)^{L}$. The variance parameters $\sigma_1,\sigma_2$ can be in general unique values- in the position eigenstate representation $\sigma_1$ controls the peak width and $\sigma_2^{-1}$ controls the width of the overall Gaussian envelope. For $\sigma_1,\sigma_2\rightarrow0$ we get to the ideal GKP qubit states. The quadrature variances are related to the peak width and envelope parameters- we shall operate with the assumption $\langle \hat{q}^2\rangle =\langle \hat{p}^2 \rangle \equiv\sigma_{q/p} =\sigma_1/2=\sigma_2/2$. The action of various channels on the quadrature variances are given by
\begin{table}[h]
	\begin{tabular}{lc}
		\toprule
		Channel & Variance Modification \\
		\midrule
		{Pure Loss }$(\eta)$& $\sigma_q^2\rightarrow\eta \sigma_q^2+(1-\eta)/2$\\
		{Phase Sensitive Amplification }$(G)$& $\sigma_q^2\rightarrow G \sigma_q^2+(G-1)/2$\\
		{Pre-Amplified Pure Loss }$(\eta;G=1/\eta)$& $\sigma_q^2\rightarrow  \sigma_q^2+(1-\eta)$\\
		Post-Amplified Pure Loss $(\eta;G=1/\eta)$&$ \sigma_q^2\rightarrow  \sigma_q^2+(1-\eta )/\eta$\\
		\bottomrule
	\end{tabular}
\end{table}
%\begin{align}
%	\text{Pure Loss }(\eta):&\, \sigma_q^2\rightarrow\eta \sigma_q^2+(1-\eta)/2\\
%	\text{Phase Sensitive Amplification }(G):&\, \sigma_q^2\rightarrow G \sigma_q^2+(G-1)/2\\
%	\text{Pre-Amplified Pure Loss }(\eta;G=1/\eta):&\, \sigma_q^2\rightarrow  \sigma_q^2+(1-\eta)\\
%	\text{Post-Amplified Pure Loss }(\eta;G=1/\eta):&\, \sigma_q^2\rightarrow  \sigma_q^2+(1-\eta )/\eta
%\end{align}

\subsection{Post-Selected Homodyne Measurements}
\label{app:gkp_measurement}
For finitely-squeezed GKP states, the logical Pauli measurements may be performed in a post-selected fashion. The outcome of the GKP $q$ quadrature measurement on the logical states $\ket{\tilde{\mathrm{L}}}; \mathrm{L}=\{0,1\}$ is a random variable 
\begin{align}
	P_{\mathrm{L}}(x)=\frac{1}{N_\mathrm{L}} \sum_{k\in\mathbb{Z}} f(x; [2k+\mathrm{L}]\sqrt{\pi},\sigma)\times \exp(-\sigma^2/2)
\end{align}
where $N_\mathrm{L}$ is the normalization constant and  $f(x; \mu,\sigma)$ is the general Gaussian distribution,
\begin{align}
	f(x; \mu,\sigma )=\frac{\exp(-(x-\mu)^2/2\sigma^2)}{\sigma \sqrt{2\pi}}.
\end{align}
Binning the measurement outcome in bins of width $\sqrt{\pi}$ i.e.$\lfloor x/\sqrt{\pi} \rfloor$ gives us an estimate of the logical state - for a particular outcome $x_0$ the parity of $\lfloor x_0/\sqrt{\pi} \rfloor $ determines the logical measurement outcome i.e.\ an even outcome signifies the logical 0 state; an odd outcome the logical 1 state.

Additional `analog information' may be gained by evaluating the fractional outcome $x_0 -\lfloor x_0/\sqrt{\pi} \rfloor  -\sqrt{\pi}/2 \equiv x_{0,f} $. The fractional outcome may be utilized to minimize the probability of a logical error in the decoding process - by choosing a rejection window of length $2v$ the outcomes may be processed as follows -- if $|x_{0,f} |\leq \sqrt{\pi}/2-v$ we use the measurement outcome $\lfloor x_0/\sqrt{\pi} \rfloor$ as the `correct' value, while if  $|x_{0,f}|> \sqrt{\pi}/2-v$ we discard the state and re-attempt the measurement with re-initialized states.

Given an outcome $x$ that falls in the post-selection window, the following probabilities can be derived --
\begin{subequations}
	\begin{align}
		\text{Probability of correctly decoding: }& P_\mathrm{correct}^{(\mathrm{L})}= \sum_{k'\in\mathbb{Z}} \int_{(2k'+\mathrm{L})\sqrt{\pi}-\sqrt{\pi}/2+v}^{(2k'+\mathrm{L})\sqrt{\pi}+\sqrt{\pi}/2-v} dx \, P_\mathrm{\mathrm{L}}(x); \\
		\text{Probability of logical error: }& P_\mathrm{flip}^{(\mathrm{L})}= \sum_{k'\in\mathbb{Z}} \int_{(2k'+\mathrm{L})\sqrt{\pi}+\sqrt{\pi}/2+v}^{(2k'+\mathrm{L})\sqrt{\pi}+3\sqrt{\pi}/2-v} dx \, P_{\mathrm{L}}(x).
	\end{align}
\end{subequations}
The probability of the measurement outcome being outside the post-selection window is hence given by,
\begin{align}
	\begin{split}
		P_\mathrm{discard}^{(\mathrm{L})}= \sum_{k'\in\mathbb{Z}} \biggl[ \int_{(2k'+\mathrm{L})\sqrt{\pi}-\sqrt{\pi}/2-v}^{(2k'+\mathrm{L})\sqrt{\pi}-\sqrt{\pi}/2+v} dx \, P_L(x) + \int_{(2k'+\mathrm{L})\sqrt{\pi}+\sqrt{\pi}/2-v}^{(2k'+\mathrm{L})\sqrt{\pi}+\sqrt{\pi}/2+v} dx \, P_L(x) 
		\biggr].
	\end{split}
\end{align}
By definition $ P_\mathrm{correct}^{(\mathrm{L})} + P_\mathrm{flip}^{(\mathrm{L})}+ P_\mathrm{discard}^{(\mathrm{L})}=1$. It has been shown~\cite{Rozpedek2023-vi} that $P_\mathrm{correct}^{(\mathrm{L})}$ and $P_\mathrm{flip}^{(\mathrm{L})}$
are tightly bounded by 
\begin{subequations}
	\begin{align}
		\int_{-\frac{1}{2} \sqrt{\pi}+v}^{\frac{1}{2} \sqrt{\pi}-v} f(x; 0,\sigma) d x \; \leq  &\;  P_\mathrm{correct}^{(\mathrm{L})}
		\; \leq \; \int_{\frac{3}{2} \sqrt{\pi}+v}^{\infty} f(x;0,\sigma) d x + 2  \int_{-\frac{1}{2} \sqrt{\pi}+v}^{\frac{1}{2} \sqrt{\pi}-v} f(x;0,\sigma) d x; \\
		2 \int_{\frac{1}{2} \sqrt{\pi}+v}^{\frac{3}{2} \sqrt{\pi}-v} f(x; 0,\sigma) d x \; \leq& \;  P_\mathrm{flip}^{(\mathrm{L})}
		\; \leq \; 2 \int_{\frac{1}{2} \sqrt{\pi}+v}^{\infty} f(x;0,\sigma) d x.
	\end{align} 
\end{subequations}
In the main text we use $P_c\equiv \int_{-\frac{1}{2} \sqrt{\pi}+v}^{\frac{1}{2} \sqrt{\pi}-v} f(x; 0,\sigma) d x$ and $P_f\equiv 2 \int_{\frac{1}{2} \sqrt{\pi}+v}^{\frac{3}{2} \sqrt{\pi}-v} f(x; 0,\sigma) d x$ instead of $P_\mathrm{correct}^{(\mathrm{L})}$ and $P_\mathrm{flip}^{(\mathrm{L})}$ respectively - hence by definition,  $P_\mathrm{discard}^{(\mathrm{L})}=1-P_c-P_f$.
	
	\section{Displacement Operation via Beamsplitter Interaction: Detailed Model}
\label{app:effective_disp_interaction}
\begin{figure}[]
	\centering
	\includegraphics[width=0.5\linewidth]{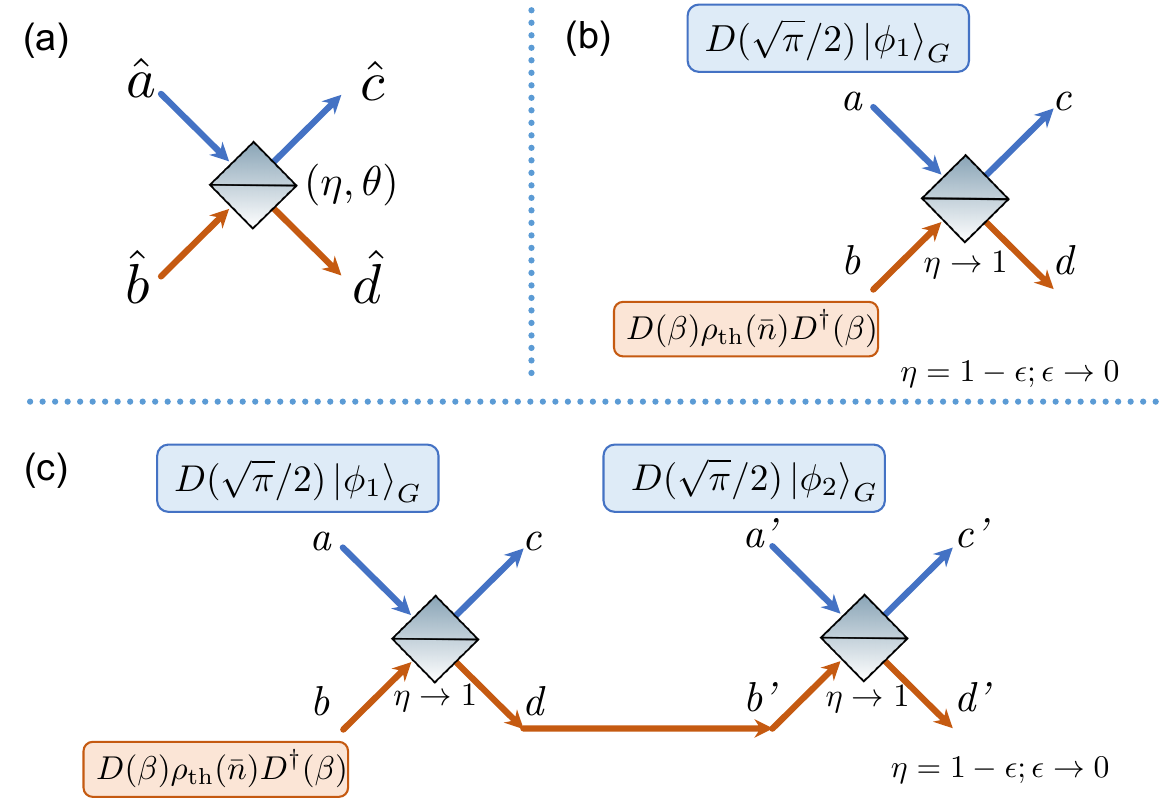}
	\caption{Beam splitter interactions -- (a) Beamsplitter convention as described in Eq.~\ref{eq:bsp_convention}; (b) Interaction of a mode $a$ carrying a displaced GKP state  }
	\label{fig:beamsplitter_int}
\end{figure}

We begin by stating the considered beamsplitter convention for input modes $\hat{a},\hat{b}$ and output modes $\hat{c},\hat{d}$ (as shown in Fig.~\ref{fig:beamsplitter_int}(a)) is 
\begin{align}
	\begin{pmatrix}
		\hat{c}\\
		\hat{d}
	\end{pmatrix}=\begin{pmatrix}
		\sqrt{\eta} & e^{i\theta}\sqrt{1-\eta}\\
		-e^{-i\theta}\sqrt{1-\eta} & \sqrt{\eta}
	\end{pmatrix} \begin{pmatrix}
		\hat{a}\\
		\hat{b}
	\end{pmatrix}.
	\label{eq:bsp_convention}
\end{align}

Analyzing this interaction using anti-normally ordered characteristic functions of the modes is quite straightforward. We  use $ \chi^{k}_A(\zeta) $ to denote the anti-normally ordered characteristic function for the mode $ k\in\{a,b,c,d\} $ w.r.t. to the (complex) variable $ \zeta $. If we are referring to the characterisitic function of a particular state, say one described by the density operator $ \rho $, we shall use the notation $ \chi^{k}_A(\zeta)[\rho] $, which is defined as,
\begin{align}
	\chi^{k}_A(\zeta)[\rho] = \Tr \left( \rho \exp(-\zeta^* \hat{a})\exp(\zeta \hat{a}^\dagger) \right) =e^{-|\zeta|^2/2} \langle D(\zeta) \rangle_\rho .
\end{align}
The characteristic function for the output modes labeled by $ c $ and $ d $ are generally given by 
\begin{subequations}
	\begin{align}
		\chi^c_{A}(\zeta)&=\chi^a_{A}(\sqrt{\eta}\,\zeta)\cdot \chi^b_{A}(\sqrt{1-\eta}\,\zeta)\label{eq:anti_normal1}, \\
		\chi^d_{A}(\zeta)&=\chi^a_{A}(-\sqrt{1-\eta}\,\zeta)\cdot \chi^b_{A}(\sqrt{\eta}\,\zeta)\label{eq:anti_normal2}.
	\end{align}
\end{subequations}

\subsection{Effective Displacement of Single Mode Input}
Consider a thermal coherent state input ${\rho}_{\mathrm{th}}(\bar{n};\beta) $ into mode $ b $ where,
\begin{align}
	{\rho}_{\mathrm{th}}(\bar{n};\beta) =D(\beta)\, {\rho}_{\mathrm{th}}(\bar{n}) \,D^\dagger(\beta).
\end{align}
Here $ \beta $ denotes the displacement applied on the thermal state $ \hat{\rho}_{\mathrm{th}}(\bar{n}) = \sum_{k=0}^{\infty} \, [\bar{n}^k/(\bar{n}+1)^{k+1}] \outprod{k}$ with a mean photon number of $ \bar{n} $. The anti-normally ordered characteristic function for this state is given by 
\begin{align}
	\begin{split}
		\chi_{A} (\zeta)[{{\rho}_{\mathrm{th}}}]=\Tr \left(  {\rho}_{\mathrm{th}}(\bar{n};\beta)  \,e^{-\zeta^* \anh{a}} e^{\zeta\crt{a}}\right)	&= \Tr \left( D(\beta)  \,  {\rho}_{\mathrm{th}}(\bar{n}) D^\dagger(\beta)  \,e^{-\zeta^* \anh{a}} e^{\zeta\crt{a}}\right)\\
		&=e^{-(1+\bar{n})|\zeta|^2} \,e^{\beta^*\zeta-\beta\zeta^* }
	\end{split}
\end{align}
Therefore as noted from Eq.~\eqref{eq:anti_normal1}-\eqref{eq:anti_normal2}, when the input mode $ b $ is a thermal coherent state, given any input mode $a$ (for e.g.\ a displaced GKP state as shown in Fig.~\ref{fig:beamsplitter_int}(b))the output state in mode $ c $ has the characteristic function
\begin{align}
	\begin{split}
		\chi^c_{A}(\zeta)&=\chi^a_{A}(\sqrt{\eta}\,\zeta)\cdot e^{-(1+\bar{n})({1-\eta})|\zeta|^2} e^{\sqrt{1-\eta}(\beta^*\zeta-\beta\zeta^*)}\\
		&={
			\chi^a_{A}(\sqrt{\eta}\,\zeta)\, e^{-({1-\eta})|\zeta|^2
			}
		}\cdot {
			e^{-\bar{n}( 1-\eta)|\zeta|^2}
		}\cdot
		{ 
			e^{\sqrt{1-\eta}(\beta^*\zeta-\beta\zeta^*)}.
		}\label{eqn:decomp_disp}
	\end{split}
\end{align}
Choose $ \beta =\beta'/\sqrt{1-\eta} $, in which case Eq.~\eqref{eqn:decomp_disp} comprises of a pure loss channel (i.e.\ $\chi^a_{A}(\sqrt{\eta}\,\zeta)\, e^{-\sqrt{1-\eta}|\zeta|^2}$), thermalizer ($e^{-\bar{n}\sqrt{1-\eta}|\zeta|^2}$) and a displacement operation ($e^{\sqrt{1-\eta}(\beta^*\zeta-\beta\zeta^*)}$) by $ \beta' =\beta $. Considering the limit where $ \eta\rightarrow1 $ (say $ \eta=1-\epsilon $ where $ \epsilon \rightarrow 0$ ) and that mode $ b $ carries a strong coherent state (i.e. the `local oscillator') satisfying $ \langle\crt{b}\anh{b}\rangle \approx| \beta'/\sqrt{1-\eta}|^2$, the mode $ c $ becomes $ \crt{c}\equiv\crt{a}+\beta' $ and mode $ d $ becomes $\crt{d}\equiv \beta' $. The output state in mode $c$ then evolves as,
\begin{align}
	\chi^c_{A}(\zeta)
	&={\chi^a_{A}(\sqrt{1-\epsilon}\,\zeta)\, e^{-{\epsilon}|\zeta|^2}}\cdot {e^{-\bar{n}\sqrt{\epsilon}|\zeta|^2}}\cdot
	{ e^{\sqrt{\epsilon}(\beta'^*\zeta-\beta'\zeta^*)}}
\end{align}

For the more general case, we show that the anti-normally ordered characteristic function for the coherent basis projector $ \proj{\alpha}{\beta} $ is given as, 
\begin{align}
	\begin{split}
		\chi^k_A(\zeta)\bigl[\proj{\alpha}{\beta}\bigr] &= \Tr\biggl[\proj{\alpha}{\beta} \,e^{-\zeta^* \anh{a}} e^{\zeta\crt{a}} \biggr]=e^{-|\zeta|^2/2} \Braket{\beta|D(\zeta)|\alpha}\\
		&=e^{-|\zeta|^2/2} e^{(\zeta \alpha^*-\zeta^*\alpha)/2}\times \exp\big[\beta^*(\alpha+\zeta)-(|\beta|^2+|\alpha+\zeta|^2)/2\big]\\
		&=\exp[-|\zeta|^2-\alpha\zeta^*+\beta^*\alpha+\beta^*\zeta-|\beta|^2/2-|\alpha|^2/2].
	\end{split}
\end{align}
When $ \alpha=\beta $, we get $ \chi^k_{A}(\zeta)= e^{-|\zeta|^2} \,e^{\alpha^*\zeta-\alpha\zeta^* }  $; when $ \alpha=-\beta $, we have $ \chi^k_{A}(\zeta)= e^{-|\zeta|^2} \,e^{-\alpha^*\zeta-\alpha\zeta^* -2|\alpha|^2}$.

\subsection{Effective Joint Displacement of Sequential Single Mode Inputs}
We may consider the action of two consecutive beamsplitter interactions on two pre-displaced GKP states as shown in Fig.~\ref{fig:beamsplitter_int}(c). We note that for the second beamsplitter the anti-normally ordered characteristic functions for the output mode $c'$ is given by $\chi^{c'}_{A}(\zeta)=\chi^{a'}_{A}(\sqrt{\eta}\,\zeta)\cdot \chi^{b'}_{A}(\sqrt{1-\eta}\,\zeta) $ where $ \chi^{b'}_{A}(\sqrt{1-\eta}\,\zeta)\equiv\chi^{d}_{A}(\sqrt{1-\eta}\,\zeta)$ i.e. the output of the first beamsplitter interaction. It is natural to note that $\chi^d_{A}(\sqrt{1-\eta}\,\zeta)=\chi^a_{A}(-({1-\eta})\,\zeta)\cdot \chi^b_{A}(\sqrt{\eta(1-\eta)}\,\zeta)$ which implies,
\begin{align}
	\begin{split}
		\chi^{c'}_{A}(\zeta)&=\chi^{a'}_{A}(\sqrt{\eta}\,\zeta)\cdot\chi^a_{A}(-({1-\eta})\,\zeta)\cdot \chi^b_{A}(\sqrt{\eta(1-\eta)}\,\zeta)\\
		&=\chi^{a'}_{A}(\sqrt{\eta}\,\zeta)\cdot\chi^a_{A}(-({1-\eta})\,\zeta){
			e^{-\eta({1-\eta})|\zeta|^2
			}
		}\cdot {
			e^{-\bar{n}\times\eta( 1-\eta)|\zeta|^2}
		}\cdot
		{ 
			e^{\sqrt{\eta(1-\eta)}(\beta^*\zeta-\beta\zeta^*)}
		}.
	\end{split}
\end{align}
Using $\eta=1-\epsilon$, we get 
\begin{align}
	\begin{split}
		\chi^{c'}_{A}(\zeta) &=\chi^{a'}_{A}(\sqrt{\eta}\,\zeta)\cdot\chi^a_{A}(-\epsilon\,\zeta){
			e^{-\epsilon({1-\epsilon})|\zeta|^2
			}
		}\cdot {
			e^{-\bar{n}\times\epsilon ( 1-\epsilon)|\zeta|^2}
		}\cdot
		{ 
			e^{\sqrt{\epsilon(1-\epsilon)}(\beta^*\zeta-\beta\zeta^*)}
		}\\
		&\approx\chi^{a'}_{A}(\sqrt{\eta}\,\zeta)\cdot
		{
			e^{-\epsilon|\zeta|^2
			}
		}\cdot {
			e^{-\bar{n}\times\epsilon|\zeta|^2}
		}\cdot
		{ 
			e^{\sqrt{\epsilon}(\beta^*\zeta-\beta\zeta^*)}
		},
	\end{split}
\end{align}
which is the effective displacement with thermalization operation on the state going in mode $a'$.

    \section{Analytic Bound for GKP State Assisted Memory Entanglement}
\label{app:analytic_bounds}
We analyze the generation of memory entanglement with the aid of GKP-qubit entanglement swap. For the purposes of this appendix, we assume that the effective $\effcnot$ gate action is perfect (i.e. no loss-induced dephasing). Starting with finite-squeezed GKP states, if the final expected state is   $\rho_{M_1 M_2} $, the final state $	\tilde{\rho}_{M_1 M_2}$ after post-selected dual-homodyne measurements is heralded with a success probability $(P_c+P_f)^2$ and is described by,
\begin{align}
	\tilde{\rho}_{M_1 M_2}=\frac{P_c^2 \, \rho_{M_1M_2}+  P_c P_f \, \rho^{(1)}_{M_1M_2} + P_f^2 \rho^{(2)}_{M_1M_2}}{(P_c+P_f)^2}, 
	\label{eq:spinspin_entapp}
\end{align}
where $\rho^{(1)}_{M_1 M_2}=\mathbb{I}_1 X_2\, \rho_{M_1 M_2}\, \mathbb{I}_1 X_2 + Z_1 \mathbb{I}_2\,  \rho_{M_1 M_2} \, Z_1 \mathbb{I}_2 $ and $\rho^{(2)}_{M_1 M_2}=Z_1X_2 \, \rho_{M_1 M_2} Z_1 X_2 $. As defined in the main text and Appendix~\ref{app:gkp_measurement}, $P_c$ is the lower bound to probability of a correctly interpreting the measurement outcome, and $P_f$ is lower bound to the probability of a logical error in the measurement. When $\rho_{M_1,M_2}$ is one of the Bell pairs $\{\ket{\Phi^\pm}_{M_1,M_2}, \ket{\Psi^\pm}_{M_1,M_2}\}$, we may use the hashing bound $I(\rho)$~\cite{Devetak2005-hi} (defined as  $I(\rho)=\max[S(\rho_A)-S(\rho_{AB}),S(\rho_B) - S(\rho_{AB})]$) to lower bound the distillable entanglement per copy generated. The hashing bound of the final state is then given by,
\begin{align}
	I(\tilde{\rho}_{M_1 M_2}) = 1+ \sum_{i=1}^4 p_i \log_2 p_i\; ,
\end{align}
where 
$p_1=  P_c^2/(P_c+P_f)^2; \; p_2=p_3=  P_c P_f/(P_c+P_f)^2; \; p_4= P_f^2/(P_c+P_f)^2$. Thus we have the overall lower bound to the distillable entanglement rate given by $\mathcal{R}(\rho_{M_1 M_2})=(P_c+P_f)^2\times I(\tilde{\rho}_{M_1 M_2})$.

To simplify the expression for $\mathcal{R}(\rho_{M_1 M_2})$, we make the substitution $P_f/P_c=x$. Thus $I(\tilde{\rho}_{M_1 M_2})$ can be simplified as,
\begin{subequations}
	\begin{align}
		I(\tilde{\rho}_{M_1 M_2})&= 1+\frac{1}{(1+x)^2}\left[x^2 \log_2\left\{\frac{x^2}{(1+x)^2}\right\}+2x \log_2\left\{\frac{x}{(1+x)^2}\right\}+\log_2\left\{\frac{1}{(1+x)^2}\right\}\right]\\
		&=1+\frac{2x(x+1)\log_2 x - 2(x+1)^2\log_2(x+1)}{(x+1)^2}\\
		&=1+ \frac{2x}{x+1}\log_2 x - 2\log_2(1+x).
	\end{align}
\end{subequations}
Consequently, the expression for $\mathcal{R}(\rho_{M_1 M_2})$ is given by 
\begin{align}
	\mathcal{R}({\rho}_{M_1 M_2})
	&=P_c^2 (1+x)^2\left[1+ \frac{2x}{x+1}\log_2 x - 2\log_2(1+x)\right].
\end{align}
It is straightforward to note that only the terms corresponding to $I(\tilde{\rho}_{M_1 M_2})$ can go to zero since both $P_c$ and $P_f$ are positive real numbers (since they correspond to probabilities). Thus the expression $\mathcal{R}({\rho}_{M_1 M_2})=0$ has the unique solution $x\approx0.123631$. Since $x$ is implicitly related to the peak variance $\sigma^2$ and the post-selection window via error-functions, analytic solutions for the threshold values of loss becomes cumbersome. The explicit expressions for $x=P_f/P_c$ is given by 
\begin{align}
	x\equiv x(\sigma,v)=\frac{
		\mathrm{erf}\left(\frac{3\sqrt{\pi}/2-v}{\sigma\sqrt{2}}\right) - \mathrm{erf}\left(\frac{\sqrt{\pi}/2+v}{\sigma\sqrt{2}}\right)
	}{
		\frac{1}{2}\left[\mathrm{erf}\left(\frac{\sqrt{\pi}/2-v}{\sigma\sqrt{2}}\right) - \mathrm{erf}\left(\frac{-\sqrt{\pi}/2+v}{\sigma\sqrt{2}}\right)\right]
	}.
\end{align}
Considering a pure loss channel of transmissivity $\eta$, the evolution of the peak variance is given by $\sigma^2\rightarrow\eta\sigma^2 + (1-\eta)$. For the infinite squeezed case, $\sigma^2\rightarrow0$, hence $\sigma_\infty (\eta)=1-\eta$. 
    
	\twocolumngrid
	
	\bibliography{biblio_spingkp_pra}
	
\end{document}